\documentstyle[pra,aps,epsf,eqsecnum]{revtex}
\draft

\author{Barbara M. Terhal$^1$ and David P. DiVincenzo$^2$}

\title{The problem of equilibration and the computation of correlation
functions on a quantum computer}

\address{\vspace*{1.2ex}
  	\hspace*{0.5ex}{$^1$ ITF, Universiteit van Amsterdam,Valckenierstraat 65, 1018 XE Amsterdam, and \\
CWI, Kruislaan 413, 1098 SJ Amsterdam, The Netherlands.}}

\address{\vspace*{1.2ex} \hspace*{0.5ex}{$^2$ IBM T.J. Watson Research
Center, Yorktown Heights, NY 10598, USA.}}

\date{\today}

\begin{document}

\pagestyle{plain}
\pagenumbering{arabic}

\maketitle

%%%%%%%%%%%%%%%%%%%%%%%%%%%%%%%%%%%%%%%%%%%%%%%%%%%%%%%%%%%%%%%%%%%%%%%%%%%%%
% Abstract

\begin{abstract}
We address the question of how a quantum computer can be used to simulate
experiments on quantum systems in thermal equilibrium. We present two
approaches for the preparation of the equilibrium state on a
quantum computer. For both approaches, we show that the output state
of the algorithm, after long enough time, is the desired equilibrium.
We present a numerical analysis of one of these approaches
for small systems. We show how equilibrium (time)-correlation
functions can be efficiently estimated on a quantum computer, given 
a preparation of the equilibrium state. The quantum algorithms that we 
present are hard to simulate on a classical computer. This indicates 
that they could provide an exponential speedup over what can be achieved 
with a classical device. 
\end{abstract}


%%%%%%%%%%%%%%%%%%%%%%%%%%%%%%%%%%%%%%%%%%%%%%%%%%%%%%%%%%%%%%%%%%%%%%%%%%%%%

% \begin{multicols}{2}[]

\def\beq{\begin{equation}}
\def\eeq{\end{equation}}
\def\bea{\begin{eqnarray}}
\def\eea{\end{eqnarray}}
\def\Adag{A^\dagger}
\def\g{\gamma}
\def\tr{\mbox{Tr }}
\def\ptr_b{\mbox{Tr }_b }
\def\ptrs{\mbox{Tr}_s }
\def\prl{\parallel}
\def\pr3{\parallel\!\!|}
\newtheorem{theo}{Theorem}
\newtheorem{defi}{Definition}
\newtheorem{lem}{Lemma}
\newtheorem{propo}{Proposition}
\newtheorem{exam}{Example}
\newtheorem{prop}{Property}
\newcommand{\dubbelR}{{\bf R}}
\newcommand{\dubbelC}{{\bf C}}

\newcommand{\vectwoc}[2]{\left(
	\begin{array}{c}{#1}\\{#2}\end{array}\right)}
\newcommand{\mattwoc}[4]{\left[
	\begin{array}{cc}{#1}&{#2}\\{#3}&{#4}\end{array}\right]}
\newcommand{\ket}[1]{\mbox{$|#1\rangle$}}
\newcommand{\bra}[1]{\mbox{$\langle #1|$}}
\newcommand{\ba}{\begin{array}}
\newcommand{\ea}{\end{array}}
\newcommand{\beitem}{\begin{itemize}}
\newcommand{\eitem}{\end{itemize}}
\newcommand{\benum}{\begin{enumerate}}
\newcommand{\enum}{\end{enumerate}}

\def\>{\rangle}
\def\<{\langle}

\def\lbL{\lb\rule{0pt}{2.4ex}}
\def\lpL{\left(\rule{0pt}{2.4ex}}
\def\lb{\left[}
\def\lp{\left(}
\def\rb{\right]}
\def\rp{\right)}

\def\ra{\rightarrow}

\def\Ybar{\bar Y}
\def\Xbar{\bar X}
\def\Zbar{\bar Z}
\def\ep{\epsilon}

\newcommand{\mypsfig}[2]{\psfig{file=#1,#2}}

\newcommand{\myfig}[4]{
	\begin{figure}[hbtp]
	\begin{center}
	\mbox{{\mypsfig{#1}{#2}}}
	\end{center}
	\caption{#3}
	\label{fig:#4}
	\end{figure}
}
%\begin{figure}[h,t,b,p]

%%%%%%%%%%%%%%%%%%%%%%%%%%%%%%%%%%%%%%%%%%%%%%%%%%%%%%%%%%%%%%%%%%%%%%%%%%%%%

\section{The limits of classical computation}
\label{theprob}

The power of quantum computers has been demonstrated in several
algorithms, of which the most striking have been Shor's factoring
algorithm \cite{shor,kitaev} and Grover's search algorithm \cite{grover}.
From the very start however, the quantum computer has
also held the promise of being a simulator of physical
systems. This is the content  of the physical version of 
the Church-Turing principle proposed by Deutsch \cite{deutsch}. 
Thus we might expect that the universal quantum
computer can be used to simulate any experiment that we could
do on a real physical system. If such a simulation can be done
efficiently (that is, without exponential slowdown), it is 
clear that this could be one of the major applications of a 
quantum computer. This promise seems to have been only partly fulfilled until now; it has been shown by several researchers \cite{lloyd_science,zalka} 
that a simulation of the unitary time evolution of a physical
system that possesses some degree of locality (which realistic
physical systems do) can be accomplished efficiently on a 
quantum computer. 
However, many quantities of interest that are determined by experiment, or by the use of classical 
simulation techniques, relate to open quantum systems, 
in particular to systems in thermal equilibrium. The thermal 
equilibrium (Gibbs) state (in the canonical ensemble) of a 
Hamiltonian $H$ is given by
\beq
\rho_{\beta}=\sum_{m=1}^N \frac{e^{-\beta E_m}}{Z} \ket{m} \bra{m},
\label{equil}
\eeq
where $\ket{m}$ ($E_m$) are the eigenvectors (eigenvalues) of $H$.
$Z$ is the partition function
\beq
Z=\sum_{m=1}^N e^{-\beta E_m},
\label{part}
\eeq
and $\beta=\frac{1}{kT}$ where $k$ is Boltzmann's constant 
and $T$ the temperature. The physical systems that concern
us in this paper will have a finite dimensional Hilbert space ${\cal H}$ that
can be decomposed as
\beq
{\cal H}={\cal H}_1 \otimes {\cal H}_2 \otimes \ldots \otimes {\cal H}_n,   
\label{decomp}
\eeq
where each ${\cal H}_i$ represents a small, constant Hilbert space, typically
associated with some (generalized) spin or other local degree of freedom.
The Hamiltonian couples these local Hilbert spaces, for example
in correspondence with a $d$-dimensional spatial lattice, so that there is
only coupling between adjacent ``spins'' on this lattice. The quantities of interest, computed in experiment or in a classical
computation, are of the form 
\beq
\tr O_1(t_1) O_2(t_2) O_3(t_3)\ldots O_k(t_k) \rho_{\beta},
\label{quant}
\eeq
where $O_i(t_i)$ are (possibly time-dependent) observables.
Both for classical systems as well as for quantum systems,
computational Monte Carlo methods have been developed to 
estimate correlation functions as in Eq. (\ref{quant}) 
\cite{hammersley,suzuki_ed,binder}. 
The quantum Monte Carlo method for systems at finite temperature
relies on a transformation introduced by Suzuki \cite{suz_paper} 
that maps an initial quantum system on a $d$-dimensional lattice onto 
a $(d+1)$-dimensional classsical system. This conversion then makes
it possible to use classical computational sampling techniques to 
estimate correlation functions as in Eq. (\ref{quant}).
There seem to be (at least) two situations when this approach runs into 
trouble and no good computational alternatives are available \cite{binder}:
(1) the correlation functions depend explicitly on time $t$, and 
(2) the quantum system is of a fermionic nature.
We will give a short explanation of why these problems are encountered.

The transformation from a classical to a quantum system is based 
on the generalized Trotter formula. Let $H=\sum_{i=1}^k H_i$ where each $H_i$ 
is a Hamiltonian on a small constant Hilbert space. The Trotter
formula reads
\beq
e^{\sigma H}=\lim_{n \rightarrow \infty} 
\left( e^{\sigma H_1/n}e^{\sigma H_2/n} \ldots e^{\sigma H_k/n} \right)^n.
\label{trotter}
\eeq
The partition function Eq. (\ref{part}) (and similarly correlation functions
as in Eq. (\ref{quant})) can be rewritten, using the Trotter formula and
the identity $\sum_m \ket{a_{i,j}^m}\bra{a_{i,j}^m}={\bf 1}$, where the pair 
of indices $(i,j)$ labels a choice of basis, as
\beq
Z=\tr e^{-\beta H}=\sum_{\{a_{i,j}\}}p_{\{a_{i,j}\}},
\eeq
where $p_{\{a_{i,j}\}}$ is a distribution over the values of the collection
of variables $\{a_{i,j}\}$ and $j$ indexes the repetitions of the factors of
Eq. (\ref{trotter}) from 1 to $n$. If the distribution is nonnegative, we can write
$p_{\{a_{i,j}\}}=e^{H_{eff}(\{a_{i,j}\})}$ where $H_{eff}$ is now a
classical Hamiltonian given by
\beq
H_{eff}(\{a_{i,j}\})=\lim_{n \rightarrow \infty} \sum_{j=1}^n 
\sum_{i=1}^k {\tilde H}_i(a_{i,j},a_{i+1,j}).
\eeq
with $a_{k+1,j}=a_{1,j+1}$, $a_{k+1,n}=a_{1,1}$ and 
\beq
{\tilde H}_i(a,b)=\log(\bra{a} \exp(-\beta H_i/n) \ket{b}).
\label{pos}
\eeq

The distribution $p_{\{a_{i,j}\}}$ will only be nonnegative when the matrix
elements $\bra{a} \exp(-\beta H_i/n) \ket{b}$ are positive. Thus it is 
important to choose the right sets of basis states $\ket{a^m_{ij}}$ to make the conversion to a
classical sampling problem with a positive distribution. There are 
fermionic systems such as certain Hubbard models \cite{binder} in
which it does not seem to be possible to choose such a good basis.
For these systems it has turned out to be very hard to get good 
estimates of correlation functions by using classical Monte Carlo
techniques. This problem is usually referred to as the ``sign'' problem.

When we are to compute time-dependent quantities, for example 
the function $f(it)=\tr e^{iHt} O_1 e^{-iHt} O_2 \rho_{\beta}$, we
need to use an imaginary time $\tau=it$ to perform the conversion of Eq. 
(\ref{trotter}) to a classical system (we expand $e^{iHt}$ with the Trotter formula).
From the classical Monte Carlo sampling 
of the function $f(\tau)$ for real $\tau$, we estimate $f(\tau)$ and 
then we could in principle analytically continue this function.
However, we only have a finite number of samples of the function 
and each sample point has some inaccuracy. The errors that are 
introduced in estimating the Fourier components $\tilde{f}(\omega)$
from this data give rise to large fluctuations when we reconstruct
$f(it)$ with the Laplace transform 
\beq
f(it)=\int_{-\infty}^{\infty} \,d\omega \,e ^{-\omega t} \tilde{f}(\omega),
\eeq
resulting in a bad approximation for the time correlation function $f(it)$.

The relevance of estimating a simple time correlation function (an example
of Eq.(\ref{quant})) such as
\begin{equation}
\tr [A(t),B(t')] \rho_{\beta}=\langle[A(t),B(t')]\rangle_s,\label{corr}
\end{equation}
where $A$ and $B$ are some Hermitian Heisenberg operators of the system, 
cannot be overestimated. Let us recall the 
many contexts in which Eq.(\ref{corr}) is used in describing 
experimental properties of many-particle quantum systems \cite{green_fun}:

When $A=B=u$, where $u$ is the displacement field of a crystal, (\ref{corr})
describes the phonon dynamics of solids as probed by inelastic neutron
scattering.  When $A$ and $B$ are the number-density operator, the dielectric
susceptibility is represented; this correlation function describes a variety
of other experiments, including x-ray photoemission and the so-called x-ray
edge singularity.  When we study the current-current response function, we
obtain the electrical conductivity as described by the Kubo formula.  (The
density-density and current-current response functions are intimately related
via the continuity equation.)  Spin-dependent quantities are also of interest:
with the spin-spin correlation function, information is obtained about the
magnetic susceptibility, and thus the magnon dynamics of ferromagnets and
antiferromagnets, the Kondo effect, and the magnetic-dipole channel in neutron
scattering.  And finally, if $A$ and $B$ involve anomalous pair amplitudes
which involve Fermion operators like $a_\downarrow(k)a_\uparrow(-k)$, the
presence and dynamics of a superconducting phase can be probed.

In short, the dynamic pair correlation functions provide a window on many
of the interesting quantities in experimental physics, and it would be
highly desirable to have a method of obtaining estimates for these quantities
by simulation on a quantum computer.  We will present some methods below
for doing this.

In this paper we develop an approach to tackle these problems on a 
quantum computer. We break the problem into two parts:
First, we present an approach to prepare our quantum computer
in the equilibrium state $\rho_{\beta}$ of a given Hamiltonian
(sections \ref{equi1} and \ref{equi2}). We will give two
alternative routes to prepare an equilibrium state.
Next we describe a procedure for efficiently estimating 
quantities as in Eq. (\ref{quant}) given that the equilibrium state 
has been prepared (section \ref{time}). 
We will not attempt to prove that our algorithms run 
in polynomial time even for a certain class of quantum systems
$H$ and/or for certain ranges of $\beta$. The equilibration problem, 
in its full generality, is expected to be a hard problem. Even classically 
there is a large class of systems that exhibit a feature called frustration, for which calculating 
the partition function $Z$ as in Eq. (\ref{part}) is 
a $P^{\sharp}$-complete problem \cite{jerrum}. Also, for these
systems, deciding whether the energy of the ground state is lower than 
some constant $K$ is an NP-complete problem \cite{barahona}.
The quantum problem has an added difficulty: We cannot assume
that we know the eigenvectors (and eigenvalues) of the Hamiltonian of
the system that we would like to equilibrate.
There has been no demonstration yet that a quantum
computer can exponentially outperform a classical computer in estimating the partition function for certain {\em classical} 
systems, which would enable us to sample efficiently from the 
classical Gibbs distribution \cite{ofer_com}.

%BMT change
The quantum algorithms that we present are hard to simulate 
on a classical computer. In both of our equilibration algorithms we 
use the fact that one can implement the unitary time evolution of a
local Hamiltonian on $n$ qubits in a polynomial number of steps in $n$ 
on a quantum computer \cite{lloyd_science}. A direct simulation of this 
procedure on a classical computer would cost exponential (in $n$) space and 
time and is therefore unrealistic. As we will show in section \ref{time}, 
given a preparation of an equilibrium state, there exists an efficient 
procedure on a quantum computer to calculate (time-dependent) correlation 
functions. As we discussed above, there is no general efficient classical 
algorithm with which one can estimate time-dependent correlation functions. 
Our quantum algorithm provides such an algorithm for a quantum 
computer. 
Lloyd and Abrams \cite{fermi} have shown that the unitary 
simulation of a fermionic system such as the Hubbard model, either in first or
second quantization, can be performed efficiently on a quantum
computer. The quantum algorithms that we will present will use this unitary 
evolution as a building block. Therefore these algorithms can be used to compute correlation functions for the Hubbard model on a 
quantum computer. This is a task for which we do not have a good 
classical algorithm, due to the ``sign'' problem, as we pointed out above.

We focus our efforts on quantum equilibration algorithms for Hamiltonians
of which the eigenvalues and eigenvectors are not known beforehand. 
These are the Hamiltonians of, for example, Heisenberg models (in more than 
two dimensions), Hubbard models, t-J models, XY models, or many-electron Hamiltonians in quantum chemistry. 
On the other hand, knowing the eigenvectors and eigenvalues of a Hamiltonian, such as in the Ising model, is no guarantee that there exists an efficient 
(polynomial time) classical algorithm that produces the equilibrium 
distribution. The situation is similar for quantum algorithms; we 
do not know in what cases the equilibration algorithms presented in 
section \ref{equi1} and \ref{equi2} give rise to a polynomial time
algorithm (see also \cite{ofer} for quantum algorithms for Ising-type
models). 

The process of equilibration is also essential in the actual 
realization of a quantum computer. One of the assumptions 
underlying the construction of a quantum computer
\cite{5rule_dpdv} is the ability to put a physical system 
initially into a known state (or a thermal equilibrium state in 
the NMR quantum computer \cite{nmr}), the computational
$\ket{00\ldots 0}\bra{00\ldots 0}$ state. The way this is done
 in an experimental setup is to let this state be the ground
state of a natural Hamiltonian and subsequently to cool to
low temperature such that the probability of being in this
ground state is some constant. This natural Hamiltonian must
be sufficiently simple for this equilibration to be achievable
efficiently and also be sufficiently weak or tunable not
to disturb the computation later on.

\section{Equilibration I}
\label{equi1}

\subsection{Introduction}
\label{intro}

The canonical ensemble is the ensemble of states
$\{p_i,\ket{\psi_i}\}$, or a density matrix
$\rho=\sum_i p_i \ket{\psi_i}\bra{\psi_i}$, such
that $\rho$ has a given energy-expectation value 
\beq
\tr H \rho=\< E\>.
\eeq
The equilibrium state in this ensemble (Eq.(\ref{equil})) can be obtained
by maximizing the von Neumann entropy of $\rho$ under this
energy constraint.
Another way in which the canonical ensemble is defined is by considering
the possible states of a system that is in contact with an
infinite heat bath at a certain temperature $T$. The total
energy of system and bath is constant, but bath and system exchange energy, so that the system equilibrates. This directly suggests that the way to prepare 
the equilibrium state on a quantum computer is to
mimic this process. In considering the computational complexity of such a procedure, we will 
have to include the space and time cost of the bath, which may be large.
Also, the intuitive picture of equilibration between a weakly coupled 
large bath and system does not tell us anything about the rate at which 
this equilibration occurs. Furthermore, the equilibration process assumes
a bath that is already in its equilibrium state. Can we make the bath
simple enough that this bath state can be prepared efficiently?
In this section we study this process of equilibration. We present 
an algorithm and we derive expressions that completely 
characterize the equilibration process in an idealized case: the
coupling between the bath and the system is very small, the bath is very large, and 
the time of interaction is large. We then proceed by a numerical 
study of the algorithm in realistic cases where the bath is of finite dimension,
the strength of the interaction is non-zero, and the interaction time
is limited.

\subsection{The algorithm}
\label{algo}

\begin{defi}{Equilibration algorithm I.} \\
Input-parameters: \\
-$H_s$, the Hamiltonian of a $N=2^n$-dimensional quantum system. \\
-$\beta$ , the inverse temperature. \\
-$H_b$, the Hamiltonian of a $K=2^k$-dimensional
``bath'' quantum system. \\
-$\lambda H_{sb}$, where $H_{sb}$ is the $NK$-dimensional ``bath-system'' 
interaction Hamiltonian and $\lambda$ is the parameter that measures the strength of the interaction between bath and system.\\
-$t$, the interaction time between bath and system.\\
-$r$, the number of times the bath is refreshed in the algorithm. \\
Define the total Hamiltonian of system and bath as
\beq
H=H_s \otimes {\bf 1}_K+{\bf 1}_N \otimes H_b +\lambda H_{sb},  
\label{def_H}
\eeq
and the trace-preserving completely positive (${\bf TCP}$) map ${\cal S}_{\lambda,t}$ as
\beq
{\cal S}_{\lambda,t}(\rho) \equiv \rm{Tr}_b \;e^{{\it i H t}} \rho \otimes \rho_{b,\beta}\,
e^{-{\it i H t}}. 
\label{def_S}
\eeq
\benum
\item {\bf Prepare system}. We prepare the $n$ qubits in the 
computational $0$ state: $\ket{000\ldots 00}\bra{000\ldots 00}$.
\item {\bf Prepare bath}. We prepare the $k$ qubits of the bath in their
equilibrium state $\rho_{b,\beta}$ of $H_b$.
\item {\bf Evolve} system and bath for time $t$ and discard bath, that is,
perform the superoperator ${\cal S}_{\lambda,t}$ of Eq. (\ref{def_S}).
\item {\bf Repeat} steps 2 and 3 $r$ times such that
\beq 
\prl {\cal S}_{\lambda,t}^{r+1}(\ket{000\ldots 00}\bra{000 \ldots 00})-
 {\cal S}_{\lambda,t}^{r}(\ket{000\ldots 00}\bra{000 \ldots 00})
\prl_{tr} \leq \epsilon,
\label{converg}
\eeq
for all $r \geq r_0$ and $\epsilon$ is some accuracy. See Appendix \ref{norms} for the definition of $\prl .\prl_{tr}$.
\enum
\end{defi}

We put several constraints on $H_s,H_b$, and $H_{sb}$. We will use local
Hilbert spaces as in Eq. (\ref{decomp}) of dimension 2 (qubits). $H_s$ must be
a ``local'' Hamiltonian. We define a $c$-local Hamiltonian on $n$ qubits
as one that can be expressed as
\beq
H_s=\sum_{i=1}^{{\rm poly}(n)} {\bf 1}_{N/c} \otimes h_i,
\label{hs_def}
\eeq 
where each $h_i$ operates on a tensor product of several small qubit
Hilbert spaces, whose total dimension is $c$.
We will also assume that the eigenvalues of $H_s$ are all 
distinct; the spectrum is non-degenerate. This will simplify the upcoming 
analysis. In order to treat Hamiltonians with degenerate spectra a change 
in the perturbation theory of Section \ref{pertu} will have to be made.  
We expect however that with that change the main result of Section \ref{find_exp}, namely succesful equilibration in the idealized case, will still hold.
$H_{sb}$ has the linear coupling form 
\beq
H_{sb}=S \otimes B,
\label{hbs_def}
\eeq
where both $S \in B({\cal H}_s)$ and $B \in B({\cal H}_b)$ are local 
Hamiltonians. $H_b$ is the Hamiltonian of a system of non-interacting qubits, i.e., it 
is a sum of single-qubit Hamiltonians:
\beq
H_b= \sum_{i=1}^{k} {\bf 1}_{K/2} \otimes h_i.
\label{hb_def}
\eeq
The bath's equilibrium state factorizes into a tensor product of qubit 
equilibrium states associated with each $h_i$:
\beq
\rho_{b,\beta}=\rho^1_{b,\beta} \otimes \ldots \otimes \rho^k_{b,\beta}. 
\eeq
This enables us to prepare the bath (step $2$) efficiently. Appendix 
\ref{bath} shows that it will cost $2k$ elementary qubit operations
to perform step $2$. The locality of $H_s,H_b$, and $H_{sb}$ is required 
in order to be able to
simulate the unitary time evolution $e^{iHt}$ in time proportional to 
$t^2/\delta$ where $\delta$ is the accuracy with which gates are implemented \cite{lloyd_science,ab_new}. 

We also choose
\beq
\< B \>_b \equiv \tr B \rho_{b,\beta}=0.
\label{bis0}
\eeq
To understand the effect of a non-zero $\< B \>_b$ we rewrite $H$ as 
\beq
H=(H_s+ \lambda \< B \>_b S) \otimes {\bf 1}_K  + {\bf 1}_N \otimes H_b+ \lambda S \otimes B', 
\eeq
where $\<B'\>_b=0$. Thus choosing a non-zero $\< B\>_b$ effectively 
corresponds to a change in the Hamiltonian of the system.
%BMT
We now discuss the last step of the algorithm, step 4.
When the superoperator ${\cal S}_{\lambda,t}$ has the equilibrium state $\rho_{s, \beta}$ as its unique fixed point, then Eq. (\ref{converg}) for all 
$r \geq r_0$ implies that 
that 
\beq
\prl {\cal S}_{\lambda,t}^r(\ket{000\ldots 00}\bra{000 \ldots 00})-
 \rho_{s,\beta} \prl_{tr} \leq \epsilon.
\eeq
for all $r \geq r_0$, that is, the equilibration process leads to 
succesful convergence to the equilibrium state. There does however not 
exist a straightforward implementation of step 4. The first problem is that we would have to 
check the closeness of the $r$th and the $(r+1)$th iteration of 
${\cal S}_{\lambda,t}$ for all $r \geq r_0$. In practice this has to be 
replaced with choosing a finite set of iterations $r$ for which the 
invariance of ${\cal S}^r(\ket{00\ldots0}\bra{00\ldots0})$ is tested. 
This problem is also encountered in classical Monte Carlo simulations.
The second problem, which is a purely quantum phenomenon, is that by 
measuring $\rho_r \equiv {\cal S}^r_{\lambda,t}(\rho)$ we might disturb 
$\rho_r$. Thus to compare $\rho_r$ with $\rho_{r+1}$ we would have to 
run ${\cal S}$ again for $r+1$ times. To assemble some statistics on 
the difference between $\rho_r$ and $\rho_{r+1}$ we have to run $r$ 
iterations of ${\cal S}$ several times. These considerations about the
verification of the convergence of the equilibration process are of course 
not special to the use of a quantum computer; they are the same as in the 
equilibration of a quantum physical system in an experimental setup.    
Furthermore, it would be an impractical task to try to measure all the 
matrix elements of $\rho_r$; $\rho_r$ contains an exponential amount 
of data of which we can extract only a polynomial amount by measurement 
in polynomial time. The best way to proceed is the same as what one 
does in classical Monte Carlo simulations \cite{binder}. If the goal 
of the computation is to estimate $\mbox{Tr } O \rho_{s,\beta}$ then 
one assembles the datapoints 
\beq
{\cal O}_r=\mbox{Tr } O \rho_r,
\eeq
until $|{\cal O}_r-{\cal O}_{r+1}| \leq \epsilon$ for a sufficiently large 
set of iterations $r \geq r_0$. The same procedure can be carried out
when the goal of the equilibration is to compute a time-dependent 
correlation function such as Eq. (\ref{quant}).

%The crucial extra difficulty of the equilibration of a quantum system compared
%to the equilibration of a classical system, lies in the fact
%that we cannot assume that we know the eigenvectors or eigenvalues of $H_s$.
%The computation of the eigenvalues to high accuracy might even on a quantum
%computer be exponentially hard, i.e. of order $poly(N)$. On a classical 
%computer the storage of the eigenvectors takes an exponential amount of space.
%On a quantum computer an efficient algorithm that finds an eigenvector, say the
%output-state is the eigenvector, still suffers from the problem of ``How do
%we get the information out in polynomial time and where do we store it?''.
%A quantum computer might have an advantage over a classical one, in the sense
%that it is possible to find approximate eigenvalues in polynomial 
%time on a quantum computer, using the essential Fourier-transform as 
%in \cite{shor,kitaev} \footnote{an insightful treatment is given in \cite{mosca}}, which was demonstrated and used in \cite{ab_new,lidar}. 
%It is, however, an open question how
%well one could estimate an eigenvalue on a classical computer, when one is
%given a polynomial description of the eigenvector and the matrix Hamiltonian
%is ``local'', which implies that it is sparse.

%BMT
In the remainder of this section we will analyse this algorithm. 
In section \ref{prop} we give some general properties of ${\bf TCP}$ maps.
In section \ref{pertu} we discuss the non-hermitian perturbation theory
that will be the basis of the analysis of ${\cal S}_{\lambda,t}$
in the idealized case. In section \ref{find_exp} we derive explicit
expressions for the idealized case. The idealized case is the case obtained 
by taking the limits $\lambda \rightarrow 0$, $k \rightarrow \infty$ and 
$t \rightarrow \infty$. We develop a perturbation theory on the basis 
of the assumption that ${\cal S}_{\lambda,t}$ of Eq. (\ref{def_S}) is 
diagonalizable. Then we can show that in this idealized case the process
has a unique fixed point which is the equilibrium state.    
Finally, in sections \ref{num} and 
\ref{num2} we present results from numerical simulations in realistic
cases. The following questions will be adressed:\\
1. How does $k$, the number of bath qubits depend on $n$, the number of
system qubits? Are they polynomially related? \\
2. What is the influence of different choices for $H_b$, $S$ and $B$ (Eqs.(\ref{hbs_def}),(\ref{hb_def}))? \\
3. How do the $r$, $\lambda$, and $t$ required for successful equilibration depend on $n$ generically? \\       

The dynamics of open quantum systems, like the system in our algorithm that 
interacts with a bath, are most often studied with the use of a 
generalized master equation \cite{fick}. The exact master equation in integral form describes the
time evolution of $\rho(t)={\cal S}_{\lambda,t}(\rho)$ of Eq. (\ref{def_S}):
\beq
\rho(t)=e^{-i {\cal L}_s t} \rho(0)-\lambda^2 \int_{0}^t\,dt' \int_{0}^{t'}\,dt''
e^{-i{\cal L}_s (t-t')} {\cal M}(t',t'') \rho(t'').
\label{exact}
\eeq
where ${\cal L}$, the Liouvillian, is defined as 
\beq
{\cal L}(\rho)=[H,\rho].
\label{defliou}
\eeq
so that ${\cal L}_s(\rho)=[H_s,\rho]$ etc. The operator 
${\cal M}(t',t'')$ is the ``memory kernel'',
\beq
{\cal M}(t',t'')=\rm{Tr}_b \,{\cal L}_{sb}\, e^{-i(1-\rho_b \rm{Tr}_b){\cal L}(t'-t'')}{\cal L}_{sb} 
\,\rho_b.
\eeq

The form in which the master equation is most often used, however, is one 
in which two simplifying approximations are made: (1) the Born 
approximation. This relates to the weakness of the interaction parameter $\lambda$. (2) the Markov approximation. The process described by 
${\cal S}_{\lambda,t}$ is Markovian if we can write
\beq
{\cal S}_{\lambda,t}({\cal S}_{\lambda,s} (\rho))={\cal S}_{\lambda, t+s} (\rho).
\label{mark}
\eeq
This typically occurs when the rate at which the effect of the system
on the bath is erased in the bath (in the sense of being spread throughout the bath) is much faster than the rate at 
which the system evolves; the system sees a ``fresh'' bath every time.
In our algorithm this loss of correlations in the bath is enforced when after some time $t$
the bath is replaced by a new bath (step $4$). We would not be able to truly
equilibrate a finite system with a finite-dimensional bath if we would
not keep refreshing it. Since the global evolution of bath and system is
unitary, eventually we will get back to the initial unentangled state and, after tracing
over the bath, to the initial state of the system (a so-called Poincar\'e recurrence).
Whether Markovian dynamics is justified will depend on the size of 
the bath, the strength of the interaction and the length of the 
interaction time. There are ways to make a simple but naive Markov approximation in 
Eq. (\ref{exact}) that lead to a master equation that fails to describe 
${\bf TCP}$ dynamics \cite{cel_loss,alicki}. The form of the master equation that 
does incorporate both the approximations and yields a physical
completely positive map is the master equation in Lindblad 
form \cite{lindblad}:
\beq
\frac{\partial \rho}{\partial t}=-i[H_s,\rho(t)]+L\rho(t)
\label{lind}
\eeq
where $L$ \cite{davies,alicki} can be expressed with a basis of operators
$F_i$ as
\beq
L\rho(t)=\frac{1}{2}\sum_{k,l=1}^{N^2-1}a_{kl} ([F_k \rho(t),F_l^{\dagger}]+
[F_k,\rho(t) F_l^{\dagger}]),
\eeq
where $a_{kl}$ is a positive semi-definite matrix. In a Lindblad equation describing the equilibration process, we expect $L$ to depend on the system Hamiltonian $H_s$. The equilibrium state $\rho_{s,\beta}$ --
if the algorithm is successful-- should be a stationary state of the process,
which implies that $[H_s,\rho_{s,\beta}]=0$ and 
\beq
L\rho_{s,\beta}=0.
\label{cond}
\eeq
Davies \cite{davies,davies1,davies2} has demonstrated that a process
described by ${\cal S}_{\lambda,t}$ where the bath is an
infinite-dimensional quantum system (for example a quantum field) {\it does} equilibrate any quantum system in the
limit where $\lambda \rightarrow 0,t \rightarrow \infty$, but 
$\lambda^2 t$ stays constant. By carefully taking a Born and Markov 
approximation, he derives a Lindblad equation of the form such that 
Eq. (\ref{cond}) is obeyed. We will perform a similar analysis here.
The main point of difference is that we use a perturbative analysis of
the dynamics which is only valid for small $\lambda^2 t$, but coincides 
in this regime with Davies' result. We furthermore obtain more
explicit expressions for the dynamics in this limit.

One can write the most general form of an $L$ that obeys a 
quantum detailed balance \cite{alicki_pap} condition, a stronger 
requirement that the stationarity of Eq. (\ref{cond}). Now, one might ask the following 
question: Could we implement this corresponding superoperator directly, 
without the use of a weakly coupled large bath, so as to save us 
time and space? We believe the answer is no, as $L$ will depend 
on the eigenvectors and eigenvalues of $H_s$, which we do not know
beforehand.

\subsection{Some useful properties of {\bf TCP} maps}
\label{prop}

In this section, we study some essential properties of the 
superoperator ${\cal S}_{\lambda,t}$ defined as in Eq. (\ref{def_S}).
This superoperator is a ${\bf TCP}$ map
\beq
{\cal S}_{\lambda,t}\,\colon \,B({\cal H}_N) \rightarrow B({\cal H}_N),
\eeq
where $B$ is the algebra of bounded operators on the Hilbert space ${\cal H}_N$. The set ${\bf TCP}[N,N]$ is the set of ${\bf TCP}$ maps ${\cal S}\colon\, B({\cal H}_N) \rightarrow B({\cal H}_N)$.

The elements of $B({\cal H}_N)$ can be represented as $N \times N$ matrices. An alternative and convenient way to represent 
$B({\cal H}_N)$ is as a $N^2$-dimensional complex vector space $\dubbelC^{N^2}$ 
\beq
I\,\colon \,\chi \in B({\cal H}_N) \rightarrow (\chi)_{ij} \in \dubbelC^{N^2}.
\eeq
This representation leads to a matrix representation of a ${\bf TCP}$ map ${\cal S}$ on $\dubbelC^{N^2}$. Let $A_i$ be the
operation elements of ${\cal S}$, i.e.
\beq
{\cal S}(\chi)=\sum_i A_i \chi A_i^{\dagger}, \;\;\; \sum_i A_i^{\dagger} A_i={\bf 1}_N.
\label{decomp_sup}
\eeq
Then  

%DPD 2/16/99 label added
%BMT

\beq
{\chi'}_{mn}=({\cal S}(\chi))_{mn}=\sum_i \sum_{k,l} (A_i)_{mk} (\chi)_{kl} (A_i^{\dagger})_{ln}
=\sum_{k,l} {\cal S}_{mn,kl}(\chi)_{kl},
\label{matrix_S}
\eeq
with
\beq
{\cal S}_{mn,kl}=\sum_i (A_i)_{mk} (A_i^{\dagger})_{ln}.
\label{reps_S}
\eeq

One can then study the eigenvectors and eigenvalues of the matrix representation of a ${\bf TCP}$ map.
First, we will give three useful properties of ${\bf TCP}$ maps
that follow directly from their definition:

\begin{prop}
Let $B_{\rm pos} \in B$ be the set of positive semi-definite matrices. Let ${\cal S} \in {\bf TCP}[N,N]$. Then
\beq
\rho \in B_{\rm pos} \Rightarrow {\cal S}(\rho) \in B_{\rm pos},
\eeq
as ${\cal S}$ is (completely) positive.
Let $\chi$ be an eigenvector of ${\cal S}$ with eigenvalue $\mu$, ${\cal S}(\chi)=\mu \chi$. We have
\beq
\tr \chi \neq 0 \Rightarrow \mu=1, 
\eeq
as ${\cal S}$ is trace-preserving.
Let $A_i$ be the operation elements in the decomposition of ${\cal S}$ as in Eq. 
(\ref{decomp_sup}). If $\chi$ is an eigenvector of ${\cal S}$ with eigenvalue 
$\mu$, then $\chi^{\dagger}$ is also an eigenvector of ${\cal S}$ with eigenvalue
$\mu^*$. This follows from 
\beq
({\cal S}(\chi))^{\dagger}=\sum_i (A_i \chi A_i^{\dagger})^{\dagger}={\cal S}(\chi^{\dagger}).
\eeq
\label{propTCP}
\end{prop}

Let $B_{{\rm pos},1}$ be the set of positive semi-definite matrices 
that have trace 1, i.e. the density matrices. Thus Property \ref{propTCP} 
implies that if a {\it density matrix} $\rho$ is an eigenvector of the superoperator, it must have eigenvalue 1, that is, 
it is a fixed point of the map. On the basis of the ${\bf TCP}$ property of a map ${\cal S}$, we can also show the
following 
\begin{propo}
Let ${\cal S} \in {\bf TCP}[N,N]$. All eigenvalues $\mu$ of ${\cal S}$ have $|\mu| \leq 1$. 
\label{mu<=1}
\end{propo}

{\it Proof} (by contradiction):
Assume $\chi$ is an eigenvector of ${\cal S}$ with eigenvalue $|\mu| > 1$.  
Note that Property \ref{propTCP} implies that $\chi$ has $\tr \chi=0$.
If $\chi$ is hermitian, $\mu$ will be real. As $\chi$ is traceless, it must have at least one negative eigenvalue. One can always find 
a density matrix $\rho$ and a small enough $\epsilon$ such that $\rho'=\rho +\epsilon \chi$ 
is still a density matrix.
Let ${\cal S}$ operate $r$ times on this density matrix.
For large enough $r$ the result  ${\cal S}^r(\rho+\epsilon \chi)={\cal S}^r(\rho)+\epsilon \mu^r \chi$ will no longer be a positive semi-definite matrix: take the eigenvector $\ket{\psi}$ of $\chi$ 
corresponding to the lowest (negative) eigenvalue $\lambda_{\rm min}$. Then 
\beq
\bra{\psi} {\cal S}^r(\rho) \ket{\psi}+\epsilon \mu^r \bra{\psi} \chi \ket{\psi} \leq 
1+\epsilon \mu^r \lambda_{\rm min},
\eeq
will become negative for large enough $r$. But Property \ref{propTCP} implies
that ${\cal S}^r(\rho')$ is a density matrix, thus $|\mu|$ cannot be larger than 1.
  When $\chi$ is non-hermitian, we reason similarly. One can find a density 
matrix $\rho$ and a small enough $\epsilon$ such that 
$\rho'=\rho+\epsilon( \chi +\chi^{\dagger})$ is a density matrix. Let 
${\cal S}(\chi)=\mu\chi=|\mu|e^{i\phi}\chi$. Let $\lambda_{\rm min,r}$ be the smallest (and negative) eigenvalue of the traceless hermitian matrix $e^{i\phi r}\chi+e^{-i\phi r}\chi^{\dagger}$. Then 
\beq
\bra{\psi} {\cal S}^r(\rho')\ket{\psi}=\bra{\psi} {\cal S}^r(\rho) \ket{\psi}
+\epsilon |\mu|^r \bra{\psi}(e^{i\phi r}\chi + e^{-i \phi r}\chi^{\dagger})\ket{\psi} \leq 1+\epsilon |\mu|^r \lambda_{\rm min,r}, 
\eeq
will become negative for some large $r$ ($\lambda_{\rm min,r}$ is a
quasi-periodic function of $r$ so it cannot be small for all large $r$).
$\Box$

Another property about the existence of fixed points can be derived:

\begin{propo}
Let ${\cal S} \in {\bf TCP}[N,N]$. ${\cal S}$ has a fixed point (which is a density matrix).
\end{propo}

{\it Proof}: 
The set of density matrices $B_{{\rm pos},1} \in B({\cal H}_N)$ 
is convex and compact. ${\cal S}$ is a 
linear continuous map and  ${\cal S}(\rho \in B_{{\rm pos},1}) \in B_{{\rm pos},1}$. Then the Markov-Kakutani Theorem V.10.6 of \cite{dun_schwartz} applies. $\Box$

%BMT
The existence of a fixed point does not by itself guarantee that the 
process described by ${\cal S}$ is ``relaxing'', that is 
$\lim_{r \rightarrow \infty} {\cal S}^r(\rho) =\rho_0$ for all 
$\rho$ where $\rho_0$ is the fixed point. The existence of such a limit 
depends on whether the fixed point is unique. 
This following Proposition proves that when there is unique fixed point, 
relaxation will occur and the relaxation rate is determined by the second 
largest eigenvalue of ${\cal S}$ \cite{priv}:

\begin{propo}
Let $\rho_0 \in B_{{\rm pos},1}({\cal H}_N)$ be the unique fixed point of 
a {\bf TCP} map ${\cal S}$. Let $|\kappa|=\max_{m| \mu_m \neq 1} |\mu_m|$, the 
absolute value of the second largest eigenvalue of ${\cal S}$. Then 
for all density matrices $\rho$ we have
\beq
\prl {\cal S}^r(\rho)-\rho_0\prl_{tr} \leq C_N {\rm poly}(r) |\kappa|^r.
\label{rate}
\eeq
where $C_N$ is a constant depending on the dimension $N$ of the system and 
${\rm poly}(r)$ denotes some polynomial in $r$. Thus for all density matrices $\rho$
\beq
\lim_{r \rightarrow \infty}\prl {\cal S}^r(\rho)-\rho_0\prl_{tr}=0.
\label{conver}
\eeq
\label{unique}
\end{propo}

{\it Proof}:
Let $\mu_i$ be the eigenvalues of ${\cal S}$. Let $s$ be the number of 
distinct eigenvalues. We can bring any matrix ${\cal S}$ into Jordan 
form $J$ by a similarity transformation $M$ \cite{horn}:
\beq
{\cal S}=M J M^{-1},
\label{similJ}
\eeq
where
\beq
J=\sum_{i=1}^s (\mu_i P_i +N_i).
\eeq
$P_i$ are orthogonal projectors and $N_i$ is a matrix 
of 1s above the diagonal in the $i$th block or $N_i$ is the 0 matrix.
When the eigenvalue $\mu_i$ is nondegenerate $N_i$ is the 0 matrix. 
We therefore have $N_iN_j=0$ for $i \neq j$ and $P_iN_j=0$ for $i \neq j$.
Call the unique largest eigenvalue $\mu_0=1$ and the corresponding projection 
$P_0$. As in Eq. (\ref{similJ}) one can write
\beq
{\cal S}^r=M J^r M^{-1}.
\eeq
where $J^r$ equals
\beq
J^r=\sum_{i=1}^s (\mu_i^r P_i+N_i')
\label{jordanr}
\eeq
where $N_i'$ is a nilpotent matrix in the $i$th block whose 
matrix elements are all smaller than or equal to $r\mu_i^r$. Note that 
$N_0$ is not present as $\mu_0$ is unique. Let ${\cal S}^0$ be $M P_0 M^{-1}$ or ${\cal S}^0(\rho)=\rho_0$.
We use $\prl A\!\prl_{tr} \leq \sqrt{N} \prl\!A\!\prl_{2}$. Note that 
$\prl\!A\!\prl_{2}$ refers to the Euclidean norm of $A$ represented 
as a vector. This follows from $(\sum_{i=1}^{N} |x_i|)^2 \leq 
N \sum_{i=1}^{N} |x_i|^2$ for complex numbers $x_i$.
We have first of all
\beq
\prl {\cal S}^r(\rho)-\rho_0\prl_{tr} \leq \sqrt{N} \prl ({\cal S}^r-{\cal S}^0)(\rho)\prl _2.
\eeq
This expression can be bounded with the use of the similarity 
transformation $M$ to 
\beq
\prl {\cal S}^r(\rho)-\rho_0 \prl_{tr} \leq \sqrt{N} \pr3 M ({\cal S}^r-{\cal S}^0) M^{-1}\pr3_2 \leq C_{1,N} \pr3 J^r-P_0 \pr3_2  
\label{simm}
\eeq
where $\pr3.\pr3_2$ is defined in Appendix \ref{norms} and we use
$\prl\!\rho\!\prl_2=\tr \rho^2 \leq 1$ for density matrices. 
Using the expression for $J^r$, Eq. (\ref{jordanr}), we can also bound
\beq
\pr3 J^r-P_0 \pr3_2 \leq  {\rm poly}(r) C_{2,N} |\kappa|^r.
\label{nilbound}
\eeq
Combining Eq. (\ref{simm}) and Eq. (\ref{nilbound}) gives us the desired result Eq. (\ref{rate}). 
Eq. (\ref{conver}) then follows as $|\kappa| < 1$ by Proposition \ref{mu<=1}.
If ${\cal S}$ is diagonalizable, the nilpotents $N_i$ in expression Eq. (\ref{jordanr}) are not present. By going through the same steps, a bound as in 
Eq. (\ref{rate}) can be derived without the factor ${\rm poly}(r)$.
$\Box$

%BMT
We refer the reader to \cite{alicki} for discussions and references 
concerning the existence of a unique fixed point and other properties 
of relaxation for a process that is described by a Lindblad equation, Eq. (\ref{lind}).

The bound on the rate of convergence of Eq. (\ref{rate}) is far from 
optimal for small $r$ as we know that for any two density matrices $\rho_1$ and $\rho_2$, 
$\prl \rho_1-\rho_2 \prl_{tr} \leq 2$. However, it is not so bad as to
invalidate the main conclusion that one would like to draw from it.
If ${\cal S}$ is diagonalizable and $|\kappa|=1-a/n^c$ then a 
(polynomial) number of iterations $r=\frac{n^c}{a}(\ln 1/\epsilon+\ln C_N)$, 
for large $n$, results in
\beq
\prl {\cal S}^r(\rho)-\rho_0\prl_{tr} \leq \epsilon,
\label{boundrate}
\eeq
where we used $\lim_{m \rightarrow \infty}(1-x/m)^m=e^{-x}$. If ${\cal S}$
is not diagonalizable the convergence is possibly slowed by the factor ${\rm poly}(r)$,
but there still will be a polynomial relation between $|\kappa|$ and $r$.

%The existence of a fixed point leaves open the question about its uniqueness.
%In the theory on irreversible processes that are described by 
%one-parameter semi-goups Davies \cite{davies} is able to define 
%properties such as recurrence, transience and irreducibility 
%that characterize classical Markov processes. We will not need such a 
%general approach here though.

%DPD 2/16/99 a new lemma; fixed 3/9/99

Finally we give a result which relates members of ${\bf TCP}[N,N]$ to the
stochastic matrices. A real matrix $M$ is stochastic when the entries 
of its columns add up to 1, i.e. $\sum_i M_{ij}=1$.

\begin{propo}
Let ${\cal S} \in {\bf TCP}[N,N]$. ${\cal S}_{mm,nn}\in\dubbelR$, and,
$\forall n$, $\sum_m{\cal S}_{mm,nn}=1$; that is, the elements ${\cal
S}_{mm,nn}$ form an $N\times N$ stochastic matrix in the indices m and n.
Also, $\forall n,k$, $n\neq k$, $\sum_m{\cal S}_{mm,nk}=0$.
\label{stochastic}
\end{propo}

{\it Proof}: 
${\cal S}_{mm,nn}\in\dubbelR$ follows directly from Eq. (\ref{reps_S}).
For the rest, we impose the unit trace condition on Eq. (\ref{matrix_S}):
\begin{equation}
1=\sum_{m,k,l}{\cal S}_{mm,kl}\rho_{kl}.
\label{sto1}
\end{equation}
This must be true for all density matrices represented by $\rho$.  Taking
$\rho_{kl}=\delta_{k,l}\delta_{k,k_0}$ gives the desired result
\begin{equation}
1=\sum_m{\cal S}_{mm,k_0k_0}.
\label{sto2}
\end{equation}
We now separate Eq. (\ref{sto1}) into diagonal and off-diagonal parts, using
the Hermiticity of the density matrix $\rho$:
\begin{equation}
1=\sum_{m,k}{\cal S}_{mm,kk}\rho_{kk}
+\sum_{m,k,l}^{k>l}({\cal S}_{mm,kl}+{\cal S}_{mm,lk}){\rm Re}(\rho_{kl})
+i\sum_{m,k,l}^{k>l}({\cal S}_{mm,kl}-{\cal S}_{mm,lk}){\rm Im}(\rho_{kl}).
\label{sto3}
\end{equation}
The first term of Eq. (\ref{sto3}) is always 1 because of Eq. (\ref{sto2}).
If we require Eq. (\ref{sto3}) when the off-diagonal terms in $\rho$ are
$\rho_{kl}=\delta_{k,k_0}\delta_{l,l_0}$ ($k>l$), we obtain
\begin{equation}
\sum_m({\cal S}_{mm,k_0l_0}+{\cal S}_{mm,l_0k_0})=0,
\end{equation}
and setting the off-diagonal terms in $\rho$ to
$\rho_{kl}=i \delta_{k,k_0}\delta_{l,l_0}$ ($k>l$) gives
\begin{equation}
\sum_m({\cal S}_{mm,k_0l_0}-{\cal S}_{mm,l_0k_0})=0,
\end{equation}
Adding these equations, we obtain the desired result
\begin{equation}
\sum_m{\cal S}_{mm,k_0l_0}=0,\ \ \ k_0\neq l_0.
\end{equation}
$\Box$

\subsection{Perturbation theory}
\label{pertu}

%BMT
In this section we develop a perturbation thoery in the coupling 
$\lambda$ for the superoperator ${\cal S}_{\lambda,t}$. The calculation
will assume the diagonalizability of ${\cal S}_{\lambda,t}$. If all the eigenvalues of a matrix $M$ are distinct, 
$M$ is diagonalizable \cite{horn}. Therefore in many cases of interest 
for equilibration, this assumption for ${\cal S}_{\lambda,t}$ will be 
correct. An example of a simple superoperator that is nondiagonalizable 
is the following. The superoperator ${\cal S}$ operates on $B({\cal H}_3)$ 
and is given by
\beq
\ba{l}
{\cal S}(\ket{i} \bra{j})=0, \mbox{      } i \neq j,  \\
{\cal S}(\ket{1} \bra{1})=\ket{2}\bra{2}, \\
{\cal S}(\ket{2}\bra{2})=\ket{2}\bra{2}, \\ 
{\cal S}(\ket{3}\bra{3})=\ket{1}\bra{1}. 
\ea
\eeq
The eigenvectors of ${\cal S}$ are $\ket{i}\bra{j}$ for all $i \neq j$, 
the state $\ket{2}\bra{2}$ and $\ket{1}\bra{1}-\ket{2}\bra{2}$. This example
shows that nondiagonalizability is not a property particular to 
superoperators describing quantum operations but is also found in 
classical Markov processes.

One can formally expand the superoperator ${\cal S}_{\lambda,t}$ as a power
series in the coupling parameter $\lambda$,  
\begin{equation}
{\cal S}_{\lambda,t}={\cal S}^{(0)}_t+\lambda {\cal S}^{(1)}_t +\lambda^2
{\cal S}^{(2)}_t +\lambda^3 {\cal S}^{(3)}_t + \ldots.\label{pert}
\end{equation}
In section \ref{find_exp} we will explicitly calculate the expressions for
these expansion operators.  We will show (Eqs. (\ref{firsts1})-(\ref{lasts1})) that condition Eq. (\ref{bis0})
implies that ${\cal S}^{(1)}_t$ is zero for all $t$.  On the basis of this
expansion, we will make a perturbative expansion of the eigenvalues and
eigenvectors of ${\cal S}_{\lambda,t}$
\begin{eqnarray}
&&\mu=\mu^{(0)}+\lambda\mu^{(1)}+\lambda^2 \mu^{(2)}+\ldots, \\
&&\chi=\chi^{(0)}+\lambda \chi^{(1)} +\lambda^2 \chi^{(2)} +\ldots.
\label{evecs}
\end{eqnarray}
Assuming that the perturbation expansion exists for this non-Hermitian
operator, it will have the same structure as in the well established procedures
familiar in quantum theory for bounded Hermitian operators (see textbooks on
quantum mechanics such as \cite{messiah} or \cite{horn} for a more
mathematical background).
%However, for non-hermitian operators such as ${\cal
%S}_{\lambda,t}$ there is no general theory that validates an expansion such as
%in Eqs. (\ref{evecs}); the $n$th-order eigenvector (eigenvalue) need not to be
%close to the $(n+1)$th-order eigenvector (eigenvalue). 
%We expect that the unitarity of the zeroth-order operator ${\cal S}^{(0)}_t$
%will permit a successful perturbative expansion of these eigenvalues and
%eigenvectors.

In the representation of Eq. (\ref{reps_S}) ${\cal S}^{(0)}_t$ reads
\begin{equation}
({\cal S}^{(0)}_t)_{mn,kl}=(U^t)_{mk} ({U^t}^{\dagger})_{ln},
\end{equation}
where $U=e^{iH_s}$. Unitarity of ${\cal S}^{(0)}_t$, as a {\em matrix}
operator on $\dubbelC^{N^2}$, follows from
\begin{equation}
\sum_{k,l} ({\cal S}^{(0)}_t)_{mn,kl}({{\cal
S}^{(0)}_t}^{\dagger})_{kl,ij}=\sum_{k,l} (U^t)_{mk} ({U^t}^{\dagger})_{ln}
(U^t)_{jl} ({U^t}^{\dagger})_{ki}=\delta_{mi} \delta_{jn}.
\end{equation}
The eigenbasis of ${\cal S}^{(0)}_t$ is formed by the set of matrices
$\rho_{nm} \equiv \ket{n} \bra{m}$ where $\ket{n}$ are the eigenvectors of
$H_s$. These eigenvectors come with eigenvalues $\mu^{(0)}_{t,nm}$:
\begin{equation}
\{\rho_{nm},\mu^{(0)}_{t,nm}=e^{it(E_n-E_m)}\}_{n,m=1}^{N,N},
\label{evals}
\end{equation}
where $E_n$ are the eigenvalues of $H_s$. Thus all density matrices of the form
$\rho_{nn}$, and mixtures of these, have degenerate eigenvalues
$\mu^{(0)}_{t,nn}=1$.  If the spectrum of $H_s$ is non-degenerate (we assumed
this in section \ref{algo}), then all other eigenvectors $\rho_{nm}$ for $n
\neq m$ have non-degenerate eigenvalues.  These eigenvectors $\rho_{nm}$ form
an orthonormal set with the vector inner product on $\dubbelC^{N^2}$,
\begin{equation}
\mbox{Tr } \rho_{nm}^{\dagger} \rho_{kl}=\delta_{nk}\delta_{ml}.
\end{equation}

%DPD 2/16/99 removed below.

%Furthermore, we note the useful fact that for the full superoperator, 
%\beq
%{\cal S}_{mm,nn} \in \dubbelR,
%\label{realD}
%\eeq 
%by using the definition in Eq. (\ref{reps_S}).

To carry out the perturbation theory, we switch to a ket notation for the
density operators and a matrix notation for the superoperators.  This will
make it easier for us to perform the necessary manipulations of degenerate
perturbation theory, in which the degenerate sector is isolated and a
diagonalization performed within it.  

We first organize the diagonal, degenerate part of this vector space to be
indexed.  To be specific, we introduce an orthogonal basis in this vector
space such that
\begin{eqnarray}
&&|\phi_i^{(0)}\rangle=\rho_{ii},\ \ 1\leq i\leq N,\\
&&|\phi_{i(m,n)}^{(0)}\rangle=\rho_{mn},\ \ 1\leq m,n\leq N,\ \ m\neq n.
\end{eqnarray}
In the second equation the indexing $i$ can be made consecutive by choosing
\begin{equation}
\begin{array}{l}
i(m,n)=nN+m-{1\over 2}n(n+1),\ \ m>n,\\
i(m,n)={1\over 2}N(N-1)+mN+n-{1\over 2}m(m+1),\ \ n>m.\end{array}
\end{equation}
This organizes this new vector space into a direct-sum form ${\bf C}^{N^2}=
{\bf C}_D \oplus {\bf C}_{N\!D}$, where ``$D$'' and ``{\em ND}'' stand for
diagonal and nondiagonal (or, degenerate and nondegenerate). ${\bf C}_D$ has
dimension $N$ and ${\bf C}_{N\!D}$ has dimension $N^2-N$.

From the discussion above, we note that the degeneracy is lifted in lowest
order by the second-order part of the superoperator ${\cal S}$ in the $D$
sector, which we will denote ${\cal S}^{(2)}_{D,D}$.  Assume that 
${\cal S}^{(2)}_{D,D}$ is diagonalizable via the similarity transformation
\begin{equation}
M{\cal S}^{(2)}_{D,D}M^{-1}=\tilde{\cal S}^{(2)}_{D,D},
\end{equation}
where $\tilde{\cal S}^{(2)}_{D,D}$ is a diagonal matrix (the tilde will denote
quantities expressed in the new basis $M_D\oplus{\bf
1}_{N\!D}|\phi^{(0)}\rangle$, which is in general non-orthogonal).  In this
new basis, the degeneracy of the diagonal terms of ${\cal S}$ is lifted to
second order in $\lambda$ (the diagonal terms can be written to second order
as $\mu_i=1+\lambda^2\tilde{\cal S}^{(2)}_{ii}$), and since the largest
off-diagonal terms in the $D$ sector are now third order, given by
\begin{equation}
\lambda^3M{\cal S}^{(3)}_{D,D}M^{-1}=\lambda^3\tilde{\cal S}^{(3)}_{D,D},
\end{equation}
the condition for the successful application of non-degenerate perturbation
theory is now satisfied, assuming that no additional, accidental degeneracy
occurs.  (The condition is satisfied from the start in the {\em ND} sector.)
Its form is essentially no different from the conventional perturbation
expansion\cite{messiah} for Hermitian operators.  This expansion for the
eigenvalues is
\begin{equation}
\mu_i=\mu^{(0)}_i+\lambda^2\tilde{\cal S}^{(2)}_{ii}+O(\lambda^3).
\end{equation}
The form of this expansion is different depending on whether $i\in D$ or $i\in
\mbox{\em ND}$, but only at $O(\lambda^4)$.  The perturbation expansions for
the eigenvectors are

%DPD 2/16/99 removed tilde from second equation.

\begin{eqnarray}
&&|\phi_i\rangle=|\tilde{\phi}^{(0)}_i\rangle+\lambda\sum_{j\in D,\ j\neq i}
|\tilde{\phi}^{(0)}_j\rangle\frac{\tilde{\cal S}^{(3)}_{ji}}
{\tilde{\cal S}^{(2)}_{ii}-\tilde{\cal S}^{(2)}_{jj}}+O(\lambda^2),
\ \ i\in D,\\
&&|\phi_i\rangle=|\phi^{(0)}_i\rangle+\lambda^2\sum_{j\neq i}
|\tilde{\phi}^{(0)}_j\rangle\frac{\tilde{\cal S}^{(2)}_{ji}}
{\mu^{(0)}_i-\mu^{(0)}_j}+O(\lambda^3),
\ \ i\in \mbox{\em ND}.
\end{eqnarray}
%DPD 2/16/99 rewriting starts here
This expansion indicates that there is no mixing between the $D$ and {\em ND}
sectors until second order in $\lambda$.  This expansion strategy will be
taken up again in the numerical simulations, Sec. \ref{num}
(Eq. (\ref{diag})).

We note that this separation of the superoperator into ${\em D}$ and ${\em
ND}$ sectors permits us to write the action of the superoperator
Eq. (\ref{pert}) using a more informative expression in which these sectors
are almost decoupled:
\begin{equation}
({\cal S}_{\lambda,t}(\rho))_{nn}=\sum_m P_{nm,t}\, \rho_{mm}+
\delta\rho_{nn},\label{marko}
\end{equation}
\begin{equation}
P_{nm,t}=\delta_{nm}
+\lambda^2
({\cal S}_t^{(2)})_{nn,mm} +\lambda^3 ({\cal S}_t^{(3)})_{nn,mm} + \ldots,
\label{Dsec}
\end{equation}
\begin{equation}
\rho_{mn}=
\lambda^2\sum_{k,l,k\neq l}({\cal S}_t^{(2)})_{nn,kl}\rho_{kl}+
\lambda^3\sum_{k,l,k\neq l}({\cal S}_t^{(3)})_{nn,kl}\rho_{kl} + \ldots.
\label{NDsec}
\end{equation}
Note from Proposition \ref{stochastic} that $P_{nm,t}$ is exactly a stochastic
matrix; therefore the dynamics in the $D$ sector is that of a classical
Markov process, up to second order in $\lambda$ (since the contribution
from the {\em ND} sector, Eq. (\ref{NDsec}), is $O(\lambda^2)$).  The
dynamics inside the {\em ND} sector is also simple:
\begin{equation}
({\cal S}_{\lambda,t}(\rho))_{nm}=\mu^{(0)}_{t,nm}\, \rho_{nm}+
\lambda^2\sum_{k,l,k\neq l}({\cal S}_t^{(2)})_{nm,kl}\rho_{kl}+\ldots,
\ \ \ n\neq m.
\label{scala}
\end{equation}
So, to $O(\lambda^2)$, there are no contributions to this equation from the
$D$ sector.  So, the low-order dynamics in the {\em ND} sector simply involves
a scalar multiplication of the off-diagonal components of the input matrix
$\rho$.

The simplications of Eqs. (\ref{marko}) and (\ref{scala}) makes it possible to
answer questions about the uniqueness of the fixed point and, in
principle, the mixing properties of a repeated application of $S_{\lambda,t}$,
using techniques from classical Markov processes \cite{markov}. The splitting
in two sectors, each having its own relaxation times, is similar to the
phenomenological description of a relaxation process by means of Bloch
equations or the Redfield equation \cite{abragam}.  This description in terms of the longitudinal relaxation time
$T_1$ ($D$ sector) and transversal relaxation time $T_2$ ({\em ND} sector) is,
for example, used in NMR \cite{abragam}.

Of course, the ``smallness'' of the operators $\lambda^2 {\cal S}^{(2)},
\lambda^3 {\cal S}^{(3)}, \ldots$ compared to ${\cal S}^{(0)}$ will determine
how fast the perturbation series converges.  We will calculate the
eigenvectors of ${\cal S}_{\lambda,t}$ to zeroth order in $\lambda$ and the
eigenvalues to second order in $\lambda$. The stochastic matrix $P_{nm,t}$ 
is determined in this approximation. The justification of this
approximation will be given when we explicitly determine the expressions for
${\cal S}_{\lambda,t}$ in section \ref{find_exp}, where we set bounds on
$\lambda$ and $t$ such that indeed $\lambda^2$ and higher order corrections
are small within some norm (for example the $\prl.\prl_{\diamond}$ given in
\cite{werner,akn}). 

%DPD 2/16/99 end of corrected section

%Thus, let ${\cal S}_{\lambda,t}={\cal
%S}^{(0)}_t+\lambda^2 {\cal S}^{(2)}_t +\sum_{n=3}^{\infty} \lambda^n {\cal
%S}^{(n)}_t$ as given in Eq.(\ref{def_S}).  We will need to show that
%\marginpar{\tiny will we do this?}
%\begin{equation}
%\prl \sum_{n=3}^{\infty} \lambda^n {\cal S}^{(n)}_t\prl_{\diamond} \ll \prl
%{\cal S}^{(2)}_t \prl_{\diamond} \ll \prl {\cal S}^{(0)}_t\prl_{\diamond}.
%\label{bound}
%\end{equation}

\subsection{Calculation of expressions}
\label{find_exp}

%DPD 3/5/99 insertion
%BMT
Here we will calculate the elements of the superoperator described in the
last section to lowest non-trivial order in $\lambda$ ($\lambda^2$).  
Truncating the expression for $P$ in Eq. (\ref{Dsec}) to second order,
$Q_{nm,t}$ is defined by the expression
\begin{equation}
P_{nm,t}\approx\delta_{nm}+\lambda^2 Q_{nm,t}.
\end{equation}
And taking $\mu$ in Eqs. (\ref{scala},\ref{NDsec}) to second order, 
and using Eq. (\ref{evals}), we define
$\nu_{nm,t}$ by
\begin{equation}
\mu_{nm,t}\approx e^{it(E_n-E_m)}(1+\lambda^2 \nu_{nm,t}).
\end{equation}
%DPD 3/5/99 end of insertion
In this section we will find expressions for $Q_{nm,t}$ and $\nu_{nm,t}$ and
exhibit the regime in which they give a valid description of ${\cal
S}_{\lambda,t}$. We also show that for a large enough bath, the equilibrium
state is the fixed point of the map $S_{\lambda,t}$. We discuss under what
conditions this fixed point is unique.

We will use operators in the Heisenberg representation. We denote 
such operators (for example on the system) as 
\beq
A_t=e^{iH_s t} A\, e^{-iH_s t}.
\eeq 
The total Liouvillian ${\cal L}$ is defined as 
\beq
e^{-i {\cal L} t}(\rho \otimes \rho_{b,\beta})= U^t (\rho \otimes \rho_{b,\beta}){U^t}^{\dagger}.
\label{liou}
\eeq
One can expand the operator $e^{-i {\cal L}t}$ in a perturbation series in $\lambda$
\cite{fick}, take a partial trace over the bath and identify the operators 
${\cal S}^{(0)}_{t}=e^{-i {\cal L}_s t}$, ${\cal S}^{(1)}_{t}$ and ${\cal S}^{(2)}_{t}$ in Eq. (\ref{pert}):
\beq
{\cal S}^{(1)}_{t}=-i\,{\rm Tr}_b \int_0^t\, dt' e^{-i({\cal L}_s+{\cal L}_b)(t-t')}
{\cal L}_{sb}\, e^{-i({\cal L}_s+{\cal L}_b)t'},
\label{firsts1}
\eeq
and 
\beq
{\cal S}^{(2)}_{t}=-{\rm Tr}_b \int_0^t dt'\, \int_0^{t'} dt''\,
e^{-i({\cal L}_s+{\cal L}_b)(t-t')}{\cal L}_{sb}\,e^{-i({\cal L}_s+{\cal L}_b)(t'-t'')}
{\cal L}_{sb}\,e^{-i({\cal L}_s+{\cal L}_b)t''}.
\label{s2}
\eeq

%BMT
First we consider ${\cal S}^{(1)}_t$. We use Eq. (\ref{liou}) and 
Eq. (\ref{defliou}) to rewrite ${\cal S}^{(1)}_t$ acting on $\rho \otimes \rho_{b,\beta}$ as:
\beq
{\cal S}^{(1)}_{t}(\rho \otimes \rho_{b,\beta})=
-i \lambda {\rm Tr}_b  \int_0^t\, dt' 
e^{i H_s (t-t')} \otimes e^{i H_b (t-t')}
\,[H_{sb}, \rho_{t'} \otimes \rho_{{b,\beta}_{t'}}]\,
e^{-i H_s (t-t')} \otimes e^{-i H_b (t-t')},
\label{s1expanded}
\eeq
where $\rho_{t'}$ is the time-evolved (with $H_s$) $\rho$ and 
$\rho_{{b,\beta}_{t'}}$ is the time-evolved (with $H_b$) $\rho_{b,\beta}$. 
The equilibrium state $\rho_{b,\beta}$ is invariant under unitary 
evolution with $e^{iH_b t'}$ and thus $\rho_{{b,\beta}_{t'}}=\rho_{b,\beta}$.
We then use the cyclic permutation invariance of the trace and 
$H_{sb}=S \otimes B$ to rewrite 
equation (\ref{s1expanded}) as a simpler sum of two terms 
\beq
{\cal S}^{(1)}_{t}(\rho \otimes \rho_{b,\beta})=
-i \lambda \int_0^t\, dt' \left[e^{i H_s (t-t')} S \rho_{t'} e^{-i H_s (t-t')}
-e^{i H_s (t-t')} \rho_{t'} S e^{-i H_s (t-t')}\right]
{\rm Tr}_b B \rho_{b,\beta}
\label{lasts1}
\eeq
Then the condition Eq. (\ref{bis0}) implies that 
${\cal S}^{(1)}_t(\rho \otimes \rho_{b,\beta})$ is 0 for any $\rho$. 
%end BMT

Let us consider the second order term. The expression for ${\cal S}^{(2)}_t$ reads
\bea
\lefteqn{{\cal S}^{(2)}_{t}=-e^{-i{\cal L}_s t} \int_0^t dt'\, \int_0^{t'} dt''\, \left( 
h(t'-t'') S_{-t'} S_{-t''} \rho  - h(t''-t') S_{-t'} \rho S_{-t''}\right.} \nonumber \\
& &  \left. - h(t'-t'') S_{-t''} \rho S_{-t'} +  h(t''-t')\rho S_{-t''} S_{-t'} \right),
\label{calc1}
\end{eqnarray}
where $h(t)$ is defined as $\<B B_t \>_b$. We write 
\beq
h(t)=\int_{-\infty}^{\infty}
d \omega \; e^{it \omega} \tilde{h}(\omega).
\label{ft}
\eeq
Let $S_{nm}$ be the matrix elements of the interaction $S$ in this eigenbasis
of $H_s$, $S_{nm}=\bra{n} S \ket{m}$. 
%BMT
Now we can find the expression for 
$Q_{mn,t}=({\cal S}_t^{(2)})_{mm,nn}$. From Eq. (\ref{calc1}) after integration over the variables $t'$ and $t''$ and with the use of Eq. (\ref{ft}), we find:
\beq
Q_{mn,t}=2\int_{-\infty}^{\infty}d\omega \; \tilde{h}(\omega)  \left[\frac{|S_{mn}|^2 (1-\cos t (\omega-E_n+E_m))} 
{(\omega-E_n+E_m)^2}-\sum_l \frac{\delta_{nm} |S_{nl}|^2 (1-\cos t (\omega-E_n+E_l))} {(\omega-E_n+E_l)^2}\right].
\eeq
%BMT
For the ``decay factor'' $\nu_{nm,t}$ in the ND sector we find 
\beq
\nu_{nm,t}=\int_{-\infty}^{\infty}\, d\omega \;
\tilde{h}(\omega) \left[\frac{2 S_{nn}S_{mm}(1-\cos t\omega)}{\omega^2}
-f(t,\omega,E_n)-f^*(t,\omega,E_m)\right],
\eeq
with $f^*$ the complex conjugate of $f$. The function $f$ is given by
\beq
\mbox{Re } f(t,\omega,E_n)=\sum_l \frac{|S_{ln}|^2 (1-\cos t (\omega-E_n+E_l))} {(\omega-E_n+E_l)^2},
\eeq
and
\beq
\mbox{Im } f(t,\omega,E_n)=\sum_l \frac{|S_{ln}|^2}{\omega-E_n+E_l}
\left[1-\frac{\sin t(\omega-E_n+E_l)}{t(\omega-E_n+E_l)}\right].
\eeq

We will now look at the idealized case, i.e., we take the limits (remember $k$ is 
the number of qubits in the bath)
\beq
P_{nm,\lambda^2 t} \equiv \lim_{\stackrel{t \rightarrow \infty, \lambda \rightarrow 0}{\tiny{\mbox{ constant }} \lambda^2 t}}\lim_{k \rightarrow \infty} P_{nm,t},\;\;\; 
\mu_{nm,\lambda^2 t} \equiv e^{it(E_n-E_m)} \lim_{\stackrel{t \rightarrow \infty, \lambda \rightarrow 0}{\tiny{\mbox{ constant }} \lambda^2 t}} \lim_{k \rightarrow \infty}
(1+\lambda^2 \nu_{nm,t}).
\eeq
When the bath is infinitely large, it will have a continuous spectrum; 
$\tilde{h}(\omega)$ will be a smooth function. The rate of interaction vanishes, but as we take the limit $t \rightarrow \infty$, there is an effective non-zero interaction 
that is proportional to $\lambda^2 t$. Recall that 
\beq
\delta(x)=\lim_{t \rightarrow \infty} \frac{1-\cos(tx)}{t \pi x^2},
\label{delta}
\eeq
where $\delta(x)$ is the Dirac delta function, which is defined as
$\int_{-\infty}^{\infty}\, dx\, \delta(x)=1$ and, $\forall\, x \neq 0$,
$\delta(x)=0$. With the use of the $\delta$ function we find
\beq
P_{mn,\lambda^2t}=\delta_{nm}(1-\lambda^2 t 2\pi \sum_l |S_{nl}|^2 \hat{h}(E_n-E_l))
+\lambda^2 t 2\pi |S_{mn}|^2  \hat{h}(E_n-E_m),
\label{def_markov}
\eeq
and  
\beq
\mu_{nm,\lambda^2 t}=e^{it(E_n-E_m)}\left(1+\lambda^2 t 2 \pi S_{nn} S_{mm} \tilde{h}(0) 
-\lambda^2 t \pi g(E_n)-\lambda^2 t \pi g^*(E_m)\right),
\label{decay}
\eeq
with 
\beq
\mbox{Re } g(E_n)=  \sum_l |S_{ln}|^2 \tilde{h}(E_n-E_l),
\eeq
and 
\beq
\mbox{Im } g(E_n)= {\cal P} \int_{-\infty}^{\infty}\,d\omega \tilde{h}(\omega) \sum_l
\frac{|S_{ln}|^2}{\omega-E_n+E_l}
\eeq
where ${\cal P}$ is the principal value of the integral.
%BMT
In order to see in what regime the perturbation theory is correct, we 
check whether the process described by Eq. (\ref{def_markov}) and 
Eq. (\ref{decay}) corresponds to that of a {\bf TCP} map. 
First we verify Property \ref{propTCP} in Eq. (\ref{decay}); the 
eigenvalues of $\ket{n}\bra{m}$ and $\ket{m}\bra{n}$ are related by complex conjugation, or $\mu_{nm,\lambda^2 t}^*=\mu_{mn,\lambda^2 t}$. 
The trace-preserving property (also in \ref{propTCP}) is 
also obeyed:
\beq
\sum_m P_{mn,\lambda^2 t}=1. 
\label{stochas}
\eeq
Complete positivity of the map implies that $P_{mn,\lambda^2 t}$ must be
a matrix of probabilities, that is we must have $P_{mn,\lambda^2 t} \geq 0$. Thus the first necessary condition for the 
validity of the perturbative approximation is 
\beq
\ba{lr}
{\bf \mbox{Condition 1:    }} & \forall\;n:\; \lambda^2 t  \ll \frac{1}{2 \pi \sum_l |S_{ln}|^2 \hat{h}(E_n-E_l)}.
\ea
\label{cond1}
\eeq
Eq. (\ref{stochas}) and Eq. (\ref{cond1}) together ensure that 
$P_{mn,\lambda^2 t}$ is a stochastic matrix. Complete positivity also 
implies via Proposition \ref{mu<=1} that $|\mu_{nm,\lambda^2 t}| \leq 1$.
In order that $|1+\lambda^2 t a| \leq 1$, where $a$ is some complex number, 
we must have that $\mbox{Re }a \leq 0$ and $\lambda^2 t \leq 2/|\mbox{Re }a|$. 
This real part in Eq.(\ref{decay}) is indeed negative as 
$\tilde{h}(\omega)$ is positive, and we obtain a new condition:
\beq\ba{lr}
{\bf \mbox{Condition 2:    }} & \forall\; m,n:\; \lambda^2 t \ll \frac{1}
{\pi |-S_{nn}S_{mm} \tilde{h}(0)+\frac{1}{2}\sum_l |S_{ln}|^2 \tilde{h}(E_n-E_l)
+\frac{1}{2}\sum_l |S_{lm}|^2 \tilde{h}(E_m-E_l)|}.
\ea
\label{cond2}
\eeq
Note that this condition is quite similar to the condition in Eq. (\ref{cond1}).

It is not hard to see that the stochastic matrix $P_{mn,\lambda^2 t}$ obeys detailed
balance for the equilibrium distribution:
\beq
P_{mn,\lambda^2 t}e^{-\beta E_n}=P_{nm,\lambda^2 t}e^{-\beta E_m},
\eeq
as the equilibrium condition of the bath implies that  
\beq
\tilde{h}(-\omega)=e^{-\beta \omega} \tilde{h}(\omega).
\eeq
Thus the equilibrium density matrix $\rho_{s, \beta}$ is a fixed point 
of the idealized equilibration process. To consider whether this
fixed point is unique, we note the following: If a stochastic matrix $M$ is such that all 
its matrix elements $M_{ij} > 0$, then $M$ has a unique eigenvalue equal to
1 \cite{hammersley}. 
If Condition 1 is obeyed, we indeed have $P_{mn,\lambda^2 t} > 0$ and therefore the absolute value of the second largest eigenvalue (in the diagonal 
sector) is smaller than 1. For the
off-diagonal sector, Condition 2 says that the largest eigenvalue in 
the off-diagonal sector is strictly smaller than 1 in absolute value.
Thus under these conditions, with Proposition \ref{unique}, we can conclude that the
process converges to the equilibrium state. The expression of $P_{mn,\lambda^2 t}$ coincides with the derivation given by Davies \cite{davies1} for small
$\lambda^2 t$.
%end BMT

One can help to speed up the process in the off-diagonal sector by 
``dephasing''; that is, after having the system and the bath interact 
for some time $t$, we perform the operation 
\beq
{\cal D}^a(\rho_s)=\frac{1}{a} \sum_{s=0}^a e^{iH_s s}\rho_s e^{-iH_s s},
\eeq 
which can be implemented with the assistance of an extra register in 
the state $\frac{1}{\sqrt{a}}\sum_{s=0}^a \ket{s}$ which is used to condition 
the evolution $U=e^{iH_s s}$ and subsequently traced out. The dephasing 
has the effect of canceling off-diagonal terms in the eigenbasis of the
system, i.e.
\beq
\lim_{a \rightarrow \infty}{\cal D}^a\left(
\sum_{k,l} \alpha_{kl} \rho_{kl}\right)=\sum_k \alpha_{kk}\rho_{kk}.
\eeq
A complete dephasing can in general not be achieved in polynomial
time in $n$ (see section \ref{equi2}), and thus must be understood as 
an extra aid but not a solution to the equilibration problem.

From the expressions for $P_{mn,\lambda^2 t}$ and 
$\mu_{nm,\lambda^2 t}$ we can understand the physical picture of the interaction
between bath and system. The system makes a transition 
from (eigen) level $n$ to level $m$, when (1) $S_{mn}$ is non-zero, (2) the bath is capable of ``receiving'' this quantum of energy 
$\Delta E=|E_m-E_n|$, that is, it has a matching energy difference 
$|\omega_i-\omega_j|=\Delta E$ {\em } and (3) $B_{ij}$ is non-zero.
Furthermore, the more such transitions there are, the faster the 
off-diagonal matrix elements decay. This confirms the 
intuitive picture that one might have of equilibration. Note also the 
similarity with the Fermi Golden Rule \cite{alicki,fetter} that describes
the transition probability from eigenlevel $n$ to $m$ in a unitary 
evolution that is perturbed by a time-dependent Hamiltonian.

For a finite-dimensional bath, we can express $h(t) \equiv \<B B_t \>$ as
\beq
h(t)=\sum_{k,l} e^{it(\omega_k-\omega_l)} |B_{kl}|^2 e^{-\beta \omega_k}/Z_b,
\label{finiteb}
\eeq
where $B_{kl}=\bra{k_b} B \ket{l_b}$ with $\ket{l_b}$ being the eigenstates of the
bath Hamiltonian $H_b$ and $Z_b$ the partition function of the bath. Taking the limits $t \rightarrow \infty$ 
and $\lambda \rightarrow 0$ before letting the bath grow large leads to 
divergent expressions for $P_{mn,\lambda^2 t}$ and $\mu_{nm,\lambda^2 t}$,
suggesting that the perturbation theory fails in this regime. This is 
not surprising, as the finiteness of the bath together with the limit 
$t \rightarrow \infty$ will lead to Poincar\'e recurrences (only the 
interaction cycle time is long due to $\lambda \rightarrow 0$).

\subsection{The inverse quantum Zeno effect}

%BMT
In our numerical studies (sections \ref{num} and \ref{num2}) we have observed a phenomenon that one might call the inverse 
quantum Zeno effect. It is a way of mapping an 
arbitrary initial state onto the completely mixed state 
${\bf 1}_N$ by interacting repeatedly and strongly with the 
state for a very short time. Here we will give a theoretical analysis that
explains this observation. Consider the weak coupling 
expansion ${\cal S}_{\lambda,t}={\cal S}^{(0)}_{t}+\lambda^2 {\cal S}^{(2)}_{t}+
{\cal O}(\lambda^3)$ with ${\cal S}^{(2)}_t$ given as in Eq. (\ref{s2}). 
We expand these operators around $t=0$:
\beq
{\cal S}_{\lambda,t}(\rho)=\rho-it[H_s,\rho]+
\frac{t^2\lambda^2}{2}([S\rho,S]+[S,\rho S])\,\<B^2\>_b+O(t^2,\lambda^3 t^3). 
\label{zeno}
\eeq
In the limit $\lambda \rightarrow \infty$, but $t \rightarrow 0$, and {\em constant} $\lambda^2 t$, the higher order terms ${\cal O}(t^2,\lambda^3 t^3)$ will vanish. Thus we see that the fixed point of 
${\cal S}_{\lambda,t}$ in this limit (assuming non-zero $\<B^2\>_b$) must obey
\beq
[H_s,\rho]=0\; \& \;[[S,\rho],S]=0.
\label{com}
\eeq
Notice that if we take the differential form of Eq. (\ref{zeno}) and the
prescribed limit, the equation is of the Lindblad form, Eq. (\ref{lind}).
 The state ${\bf 1}_N$ 
certainly meets the requirements of Eq. (\ref{com}), but is it unique? If $S$ and $H_s$ are such that they have no eigenspaces (except for the full space) in 
common, and both have a non-degenerate spectrum, we can show that ${\bf 1}_N$
is the unique eigenvector. Eq. (\ref{com}) requires that either $[S,\rho]=0$ or $[S,\rho]$ 
is diagonal in the same basis as $S$. If $[S,\rho]=0$ but also $[H_s,\rho]=0$, then $\rho$ can only be the state ${\bf 1}_N$. What happens if $[S,\rho]$ is
just diagonal in the same basis as $S$? Let $\ket{n}$ be an eigenvector of $S$ with eigenvalue
$\lambda_n$. We have for $n \neq m$
\beq
\bra{n}\,[S,\rho]\,\ket{m}=0.
\eeq
Rewriting this expression gives 
\beq
\forall n,m,\; n\neq m\;\; \bra{n}\, \rho \, \ket{m} (\lambda_n-\lambda_m)=0.
\eeq
Now, because $\rho$ is diagonal in the basis of $H_s$ as $[H_s,\rho]=0$ and 
$H_s$ and $S$ have no eigenvectors in common, there exist $n$ and $m$ such that $\bra{n} \rho \ket{m} \neq 0$. But the eigenvalues of
$S$ were non-degenerate, thus we obtain a contradiction. $\Box$

When ${\bf 1}_N$ is the unique eigenvector of this process, then, with 
the use of Proposition \ref{unique}, the repeated application as in 
step $4$ of the {\it Equilibration algorithm} I will eventually bring 
the system to the state ${\bf 1}_N$.
 
We showed that for this ``inverse quantum Zeno'' effect to occur $S$ and 
$H_s$ have to be such that they have no partial eigenspace in common and 
both have a non-degenerate spectrum. If we assume that $S$ and 
$H_s$ are $c$-local with $c$ larger or equal to 4, then this does not impose 
a very strong constraint on $S$ and $H_s$; the effect will 
occur for a generic $S$ and $H_s$.

\subsection{Specifications of the numerical simulation}
\label{num}

The main purpose of this study is to understand the effects of bath 
size and the choice of bath and interaction Hamiltonians for 
a specific system Hamiltonian. In Table \ref{settings} we list 
some of the choices that have been made
in the numerical analysis. We have randomly generated the elementary 
Hamiltonians $h_i$ that make up $H_s,H_b$ and $H_{sb}$, Eqs.(\ref{hs_def}),
(\ref{hbs_def}), (\ref{hb_def}), with a measure 
${\cal M}$. 
%BMT
We choose the diagonal elements of each $h_i$ uniformly in 
$[-a,a]$, where $a$ is sampling scale in Table \ref{settings}. The 
absolute value of the above-the-diagonal elements of a $h_i$ are chosen 
uniformly in $[0,a]$ and its phase is chosen uniformly in $[0,2\pi]$. 
The below-the-diagonal elements of $h_i$ follow from Hermiticity.
This defines ${\cal M}$. Note that ${\cal M} $ is not a unitarily 
invariant measure.

We take the Hamiltonians $S$ and $B$ as sums of all possible 
local 2-qubit interactions ($c_s=4$ in Table \ref{settings}). For the 
Hamiltonian of the system $H_s$ we also take a sum of all 
possible local 2-qubit interactions. Note that this includes a set 
of Hamiltonians that exhibit frustration, for which we don't 
expect equilibration to be particularly fast. 

In section \ref{find_exp} we observed that matching energy differences between
bath and system are an important ingredient in the equilibration of the system, which is consistent with the intuitive picture of equilibration that was 
sketched in section \ref{intro}. However, as we do not know the eigenvalues of the system, we can only pick our bath so as to optimize the 
chance for matching level differences. The sampling scale of the 
bath $f(n,k,c_s,c_b)$ is determined by roughly optimizing
these coincidences, $\Delta E_b=\Delta E_s$. 

Consider the density of states $p_s(E,a_s)$ of the system (the distribution
of eigenvalues generated by the measure ${\cal M}$) and the density of 
states $p_b(E,a_b)$ of the bath. Here $a_s$ is the sampling
scale of the system which we set to 1 (see Table \ref{settings}).
The quantity $[{\rm Tr} H_s]_{\cal M}$ is the mean and $\frac{[{\rm Tr} H_s^2]_{\cal M}}{N}$ is
the variance of the distribution $p_s(E,a_s)$.
The choice for ${\cal M}$ ensures that the distributions are symmetric 
around $E=0$: 
\beq
[\tr H_s]_{\cal M}=[\tr H_b]_{\cal M}=0.
\eeq
To optimize for matching we choose the variances to be equal:
\beq
\frac{[\tr H_s^2]_{\cal M}}{N}=\frac{[\tr H_b^2]_{\cal M}}{K}.
\label{match}
\eeq 
For large $K$ the bath distribution will be Gaussian 
(central limit theorem), whereas the system distribution will 
be similar to a Gaussian distribution for large $N$ (see Fig. \ref{dens}). 
Thus, setting the variances equal brings the distributions close together.

Consider first $[\tr H_b^2]_{\cal M}$. It is straightforward to calculate the variance of the eigenvalues of a qubit bath. Given a $2 \times 2$ Hermitian 
matrix $m_{ij}$, the eigenvalues $e_{\pm}=\frac{1}{2}(m_{11}+m_{22}\pm \sqrt{(m_{11}-m_{22})^2+4|m_{12}|^2})$ have  
\beq
[e_{\pm}^2]_{\cal M}=\frac{1}{4a_b^3}\int_{-a_b}^{a_b} \,dm_{11}\,\int_{-a_b}^{a_b} \,dm_{22}\, \int_{0}^{a_b} \, d|m_{12}|\,  e_{\pm}^2=\frac{2 a_b^2}{3}.
\eeq
Let $v_i$ be some $\pm$ pattern $i$ of length $k$, corresponding to selecting
$e_{+}$ or $e_{-}$ for each qubit bath. Let $E_{v_i}$ be an 
eigenvalue of the full bath, i.e.,
 $E_{v_i}=\sum_{m=1}^{k} e_{v_i[m]}$ where $v_i[m]$ indicates that we select
the $m$th bit in $v_i$. Then 
\beq
\frac{[\tr H_b^2]_{\cal M}}{K}=\frac{1}{K}\sum_{i=1}^K [E_{v_i}^2]_{\cal M}=\frac{2 k a_b^2}{3}.
\label{var_b}
\eeq
We calculate $[\tr H_s^2]_{\cal M}=\sum_{i,j} [|(H_s)_{ij}|^2]_{\cal M}$ for 
$n > 2$. We can write
\beq
\sum_{i,j} [|(H_s)_{ij}|^2]_{\cal M}=\sum_{i,j} \sum_{m=1}^{{n \choose 2}}
[|(h_{m})_{ij}|^2]_{\cal M}, 
\eeq
where $h_{m}$ is the $m$th local interaction Hamiltonian. We have used $[(h_{k}^*)_{ij} (h_{m})_{ij}]_{\cal M}=0$. Each row of $h_{m}$ has 
only four non-zero entries as the dimension of the local Hamiltonians $c_s$
was set to four. Using the fact that 
$[|(h_{m})_{ij}|^2]_{\cal M}=\frac{1}{3}$ for all interaction terms $m$, 
we obtain
\beq
\frac{[\tr H_s^2]_{\cal M}}{N}=\frac{4}{3} \left(\ba{c} n \\ 2 \ea\right).
\label{var_s}
\eeq
For $n=1$, we have $\frac{[{\rm Tr} H_s^2]_{\cal M}}{N}=\frac{2}{3}$. Comparing Eqs. (\ref{var_b}) and (\ref{var_s}) gives the expression for $a_b$:
\beq
a_b=f(n,k,4,2)=\sqrt{\frac{2}{k}  \left(\ba{c} n \\ 2 \ea\right)}.
\label{ss}
\eeq
For $n=1$, $f(1,k,4,2)=\sqrt{1/k}$. Fig. \ref{dens} illustrates how this setting determines the density of states of bath and system.

The numerical work consists of a calculation of the fixed point of 
${\cal S}_{\lambda,t}$ as a function of $t$ for a fixed $\lambda$ and 
the second largest eigenvalue for different baths 
and different systems and temperatures. We follow a numerical procedure
based on perturbation theory 
(Section \ref{pertu}) to perform a stable numerical evaluation of these quantities. 
%BMT
We can trust the answers from the numerical procedure only if we are 
in the regime in which perturbation theory is correct. This regime was 
heralded by the two conditions Eq. (\ref{cond1}) and Eq. (\ref{cond2})
in section \ref{equi1}. Whether these conditions are obeyed depends on the 
specific choices of $H_s$, $H_b$ and $S$ and $B$. We prefer to reformulate 
these conditions here such that they are obeyed for the average 
bath, system and interaction Hamiltonian obtained by sampling using ${\cal M}$ and the sampling scale. As the conditions are very similar, we take 
the first one,  Eq. (\ref{cond1}), and reformulate it as 
\beq
c(t) \equiv \lambda^2 t \,2 \pi \frac{NK [S^2]_{\cal M} [B^2]_{\cal M}}{W_b} \leq 1.
\label{approx_c}
\eeq
where $[S^2]_{\cal M}$, the average matrix element, is defined as 
\beq
[S^2]_{\cal M}=\frac{1}{N^2} \sum_{i,j} [|S_{ij}|^2]_{\cal M}=\frac{1}{N^2} 
[\mbox{Tr}_s S^2]_{\cal M},
\eeq
and similarly for $[B^2]_{\cal M}$. $W_b$ is the spectral width of the 
bath, i.e., 
\beq
W_b^2=\frac{[\tr H_b^2]_{\cal M}}{K}.
\label{width}
\eeq
%DDV 3/8/99
Here we indicate the approximations made in obtaining Eq. (\ref{approx_c}) from
Condition 1 (Eq. (\ref{cond1})):
\beq
\lambda^2 t \  2 \pi \sum_l |S_{ln}|^2 \hat{h}(E_n-E_l)\ll 1.
\label{cond1x}
\eeq
Using Eq. (\ref{finiteb}) and Eq. (\ref{ft}) we write the $\hat h$ function
as
\beq
\hat{h}(E_n-E_l)=\sum_{k,m} \delta((E_n-E_l)-(\omega_k-\omega_m))
|B_{km}|^2 e^{-\beta \omega_k}/Z.\label{hexpr}
\eeq 
We will approximate the matrix elements $|B_{kl}|^2$ as constants and replace
them by their average $[B^2]_{\cal M}$.  Then we can use 
density-of-states arguments to approximate the $m$ sum over the $\delta$
functions by the inverse of the average spacing between the $\delta$
functions; this spacing is given by $W_b/K$:
\beq
\sum_m \delta((E_n-E_l)-(\omega_k-\omega_m))\approx {K\over W_b}.
\eeq 
With these approximations, the partition-function sum over 
$k$ in Eq. (\ref{hexpr}) becomes exactly one.  So, Eq. (\ref{hexpr}) becomes
\beq
\hat{h}(E_n-E_l)\approx{K[B^2]_{\cal M}\over W_b}.\label{hexpr2}
\eeq 
Now Eq. (\ref{cond1x}) is
\beq
\lambda^2 t \  2 \pi {K[B^2]_{\cal M}\over W_b}\sum_l |S_{ln}|^2 \ll 1.
\label{cond1b}
\eeq
If we again approximate the matrix elements $|S_{ln}|^2$ as constants and
replace them by their average $[S^2]_{\cal M}$, and note that the
$l$ sum in Eq. (\ref{cond1b}) has $N$ terms, we obtain Eq. (\ref{approx_c}).

For the simulations we have performed, we
% end DDV 3/8/99
can find the values for $[S^2]_{\cal M}$ and $[B^2]_{\cal M}$ (note that
these Hamiltonians have locality parameter $c=4$, as does the system Hamiltonian $H_s$) and 
obtain the expression 
\beq
c(t)= \lambda^2 t \, \frac{16 \pi}{3 \sqrt{3}}  
\left(\ba{c} k \\ 2 \ea\right) \sqrt{\left(\ba{c} n \\ 2 \ea\right)}  \ll 1.
\label{valid}
\eeq
for $n > 1$ and $k > 1$. For a qubit system, $n=1$, and $k > 1$ we obtain  
\beq
c_1(t) \equiv \lambda^2 t \frac{8 \pi \sqrt{2}}{3 \sqrt{3}}  
\left(\ba{c} k \\ 2 \ea\right)  \ll 1.
\eeq

%BMT
The quantity $c(t)$ in Eq. (\ref{approx_c}) will function as a rescaled 
time which depends on the strength of $\lambda$ and the size of system 
and bath. In the regime where $c(t) \leq 1$ we expect a pertubative 
calculation of the eigenvectors and eigenvalues of the superoperator 
to be fairly accurate. 
The dimensionless parameter associated with the temperature is 
given by
\beq
\beta'=\beta \, W_s,
\eeq
where $W_s$ is the spectral width of the system, Eq. (\ref{width}) ($W_s=W_b$).
From here on, $\beta$ will refer to this scaled dimensionless parameter.
Instead of expanding the superoperator 
${\cal S}$ in a series in $\lambda$ as in Eq. (\ref{pert}), we write

%DPD 2/16/99 label added here BMT

\beq
\lambda^2 \bar{{\cal S}}^{(2)}_t \equiv {\cal S}_{\lambda,t}-{\cal S}^{(0)}_t,
\label{diag}
\eeq
where all higher order terms are grouped in $\bar{{\cal S}}^{(2)}_t$. The 
calculation of eigenvalues and eigenvectors then follows the analysis
of Section \ref{pertu}.
We find that the choice for the bath and the interaction Hamiltonian
influences whether the equilibration will succeed or not. Let 
\beq
{\cal D} \equiv \prl \rho_{s,\beta}-\rho_{0} \prl_{tr},
\label{defdist}
\eeq
where $\rho_0$ is the unit eigenvector obtained from the numerics. 
In Figs. \ref{fig_yes} and \ref{fig_no} two extrema in dynamics are shown, 
each corresponding to a different choice for the system, 
bath, and interaction. In Fig. \ref{fig_yes} the equilibration is successful, 
whereas in Fig. \ref{fig_no} the equilibration fails. $\mbox{R}_{D}$ is defined as 
\beq
R_{D}=\frac{1-|\kappa_D|}{\bar{c}(t)},
\label{scaled_r1}
\eeq
where $\kappa_D$ is the second largest eigenvalue in the diagonal sector and 
$\bar{c}(t)$ is the average coupling strength in the time interval that we
consider, which is $c(t) \in [0,0.3]$ here. Similarly, we define
\beq
R_{ND}=\frac{1-|\kappa_{ND}|}{\bar{c}(t)}
\label{scaled_r2}
\eeq
for the nondiagonal sector. 

\subsection{Numerical results for equilibration}
\label{num2}

We are interested in how well a randomly chosen bath and interaction 
equilibrate a system and how these averages are improved by choosing
larger baths. As the mixing rates and the distance to the equilibrium 
state will in general be oscillating functions of the scaled time 
$c(t)$ (see Fig. \ref{fig_no}) we will compute time averaged rates over a reasonable interval
in $c(t)$, 
\beq
[c(t_{init})=0,c(t_{end})=0.5],
\label{c_interval}
\eeq
such that we are in the realm where perturbation theory is valid, Eq. (\ref{approx_c}). We denoted these time averages (not to be confused with bath averages) as $\overline{R}_D$ and $\overline{{\cal D}}$ for the time averaged trace
distance, Eq. (\ref{defdist}), etc.
In Fig. \ref{histo} we present histograms that show how, for a given fixed
system {\it and} interaction, the equilibration process is different for
a set of randomly chosen baths with fixed dimension. The insets show the 
distribution for the lowest bin. 
%BMT
The vertical axis denotes the percentage 
of baths (the interval $[0\%,100\%]$ is given as the interval $[0,1]$) for a 
certain distance and rate. We observe that the diagonal rate 
distribution is very broad, and therefore the mean of the distribution is
not a very good (or a very stable) measure of the generic behavior. Furthermore,
we find that the rate in the diagonal sector is much worse than in the 
nondiagonal sector and thus is the dominant factor in setting the mixing time. 
This conforms to the pattern in many quantum systems, for example for nuclear
spins as observed by NMR, for which $T_1$ is generically larger than $T_2$ \cite{abragam}. 

To study the dependence on $\beta$ and on the dimension of the bath versus 
the dimension of the system, we compute the following data. We pick 
a system Hamiltonian $H_s$ of $n$ qubits that has some well 
spread out spectrum. We set the dimension of the bath and then we randomly
pick both the bath Hamiltonian and the interaction Hamiltonian. Means are denoted as $[.]_{{\cal M}b}$. For the rates we look both at the mean and 
the median. The median is denoted as $[[.]]_{{\cal M}b}$, see Fig. \ref{dim2}. The results for $n=1,2,3$ and $4$ are shown in Figs. \ref{dim2}-\ref{dim16}. We have given the median when the mean does not give a good 
representation of the distribution.  

These data clearly indicate that larger baths improve the process of 
equilibration, both in the rates (D and ND) as well as in the closeness to the
equilibrium state. The effects are the most pronounced at low temperature, 
where equilibration is in general harder as the system must relax to 
a single pure ground state. To understand the closeness scale, we show in 
Appendix \ref{norms} how far apart two arbitrarily chosen density matrices are; this
number lies around 1 for the dimensions that were considered. For these 
estimates, we see a trend towards approximations getting worse for larger
system sizes for low temperature. The scaled rates $[\overline{R_{D}}]_{{\cal M}_b}$ and $[\overline{R_{ND}}]_{{\cal M}_b}$ seem to be 
fairly constant, thus we see behavior that suggests that the rates are 
polynomially related to both system and bath number of qubits. We also observe that the nondiagonal rate (ND) is 
always higher than the diagonal rate (D).
The data show a system Hamiltonian dependence, that is, the average 
equilibration for $n=4$ seems to be more succesful than for $n=3$. 
We also observe that the difference between $T_1$ and $T_2$ becomes 
smaller with increasing $\beta$ (lower temperature).
Thus, in conclusion, it seems if we pick a bath size (in number of qubits) that is polynomially
related to the system size (note that the number of eigenvalues is then 
{\it exponentially} related), the rates of relaxation are polynomially related 
to the system size (in qubits); however the relaxed state could be still
fairly far away from the true equilibrium state for large system sizes.

\section{Equilibration II}
\label{equi2}

We present an alternative to the algorithm in section \ref{equi1}. This 
algorithm relies on the technique for the estimation of eigenvalues, originally given in \cite{kitaev} (see \cite{mosca,jozsa}). This eigenvalue estimation 
routine has also been used as a building block in an interesting 
quantum algorithm in \cite{ab_new} and \cite{lidarwang}.

Let $H_s$ be the $c$-local Hamiltonian with non-degenerate eigenvalues as 
in section \ref{equi1}. Order the eigenvalues as 
$E_0 > E_1 > \ldots > E_N$.

\begin{defi}{Equilibration algorithm II.}
\benum
\item {\bf Initialize} the system in the (infinite temperature) 
completely mixed state ${\bf 1}_N$.  Also add one $m$-qubit register set 
to $\ket{00\ldots 00}\bra{00 \ldots 00}$.
\item {\bf Compute eigenvalues} with the use of the Fourier transform 
and {\bf dephase} in computational eigenvalue basis, which will result in state
\beq
\sum_{n=0}^{N-1}\sum_{s=1}^{2^m-1} p(s,n) \ket{n} \bra{n} \otimes 
\ket{s} \bra{s},
\eeq
where $p(s,n)$ is a distribution, peaked at $s \sim E_n$ for large $m$.
The dephasing is a simple superoperator ${\cal D}$ on the eigenvalue register
that operates as 
\beq
\ba{lr}
{\cal D}(\ket{s_i}\bra{s_i})=\ket{s_i} \bra{s_i}, & {\cal D}(\ket{s_i}\bra{s_j})=0.
\ea
\eeq
\item {\bf Prepare} an additional $N$-dimensional quantum system,
the bath, also in ${\bf 1}_N$. Add a $m$-qubit register
and one qubit register set to $\ket{00\ldots 00}\bra{00 \ldots 00}$.
\item {\bf Compute eigenvalues} of the bath as for the system in step $2$.
\item {\bf Interact} system and bath according the following rule ${\cal R}$
(``partial swap''):
\beq
U_{\cal R} \ket{n,m}\ket{s,t}\ket{0}=
\left\{
\ba{lr} 
\ket{m,n}\ket{s,t}\ket{0} &
\mbox{if } t < s   \\
(p_{st}^{\beta/2}\ket{t,s}\ket{0}+\sqrt{1-p_{st}^{\beta}}
\ket{s,t}\ket{1})\ket{s,t} & \mbox{if } t \geq s 
\ea\right.
\eeq 
where $p_{st}^{\beta}=e^{-\beta(t-s)}$. 
%The circuit is shown 
%is Figure (\ref{rule}).
\item {\bf Trace} over the single-qubit register, all bath registers, and
the eigenvalue register of the system. The system will be in some state
\beq
\rho_s=\sum_n \alpha_n \ket{n} \bra{n}.
\eeq
The steps 2-6 define a ${\bf TCP}$ map ${\cal S}$, 
${\cal S}({\bf 1}_N)=\rho_s$. 
\item {\bf Repeat} steps 2-6 $r$ times such that
\beq 
\prl {\cal S}^{r+1}(\ket{000\ldots 00}\bra{000 \ldots 00})-
 {\cal S}^{r}(\ket{000\ldots 00}\bra{000 \ldots 00})
\prl_{tr} \leq \epsilon,
\label{converg2}
\eeq
for all $r \geq r_0$ and $\epsilon$ is some accuracy.
\end{enumerate}
\end{defi}

The advantage of this algorithm is its simplicity and its similarity to
a classical algorithm; we create a Markov chain in the eigenbasis of the
system. The disadavantage of the algorithm is that it is very likely 
to be slow; the computation of the eigenvalues to high accuracy 
with the use of the Fourier transfrom is very likely to be exponential in the number of qubits of the system and has to be performed twice, for system and bath, in each round of the  
chain. First, let us show that in the case when the eigenvalues are computed exactly in steps $2$ and $4$, i.e, $p(s,n)=\delta_{E_n,s}/N$ the Markov chain equilibrates the system.
Recall \cite{mosca} that the routines of steps $2$ and $4$ compute 
rescaled eigenvalues
\beq
E'_n=f_1 E_n +f_2,
\eeq
with $f_1$ and $f_2$ depending on the maximum and minimum eigenvalue (of which 
we assume that we can find an estimate) such that $E'_n \in [0,2 \pi)$. In 
the following we will drop these primes. The chain that is created can be represented as
\beq
\sum_n \alpha_n^{(k)} \ket{n} \bra{n},
\eeq
where $\alpha_m^{(k)}= \sum_n \alpha_n^{(k-1)}P_{n \rightarrow m}$.
We have
\beq
P_{n \rightarrow m}=\left\{\ba{lr} \frac{1}{N} & \mbox{if } E_m < E_n  \\
\frac{1}{N}(1+\sum_{k \leq n}(1-p_{nk}^{\beta})) & \mbox{if }E_m=E_n  \\
\frac{1}{N}p_{nm}^{\beta} & \mbox{if } E_m > E_n 
\ea
\right.
\eeq 
Note that $\sum_m P_{n \rightarrow m}=1$ as required. The equilibrium 
state Eq. (\ref{equil}) obeys the detailed balance condition: 
\beq
\forall n,m \;\;P_{n \rightarrow m}e^{-\beta E_n}=
P_{m \rightarrow n}e^{-\beta E_m}.
\eeq
 
All the matrix elements of the Markov matrix $P_{n \rightarrow m}$ are 
nonzero. Therefore the chain will have a unique fixed point which 
is equal to the equilibrium state due to detailed balance. Thus for 
all probability distributions $\alpha_n$ we have
\beq
 \lim_{k \rightarrow \infty} \sum_n \alpha_n P_{n \rightarrow m}^{(k)}=
\frac{e^{-\beta E_m}}{Z}.
\eeq

Notice that it is not hard to prepare the initial states of system and 
bath. One way to make the completely mixed state ${\bf 1}_N$ is to
make a maximally entangled state $\frac{1}{\sqrt{N}}
\sum_{i=0}^{N-1}\ket{i} \ket{i}$ and trace over the
second register. This takes $O(n)$ steps. The partial swap in step $5$
can be implemented with $O(n)$ elementary qubit steps. The dephasing in 
step $2$ is introduced to keep the form of the 
algorithm clean, but it does not affect its output. This dephasing is implemented
by measuring the eigenvalue register in the computational basis and 
discarding its answer.
When using an $m$-bit eigenvalue register the joint probability 
$p(n,s)$ in the first round (after step $2$) is equal to
\beq
p(n,s)=\frac{1}{N} \left|\frac{1}{2^m}\sum_{l=0}^{2^m-1} e^{i l(E_n-2 \pi s/2^m)}\right|^2.
\eeq
When $p(n,s)$ is not a delta function on the eigenvalue, the Markov chain 
will still be in the eigenbasis of the system; It will be a concatenation 
of chains; the transition probability of this 
new chain is
\beq
P_{n \rightarrow m}'=\sum_{s,t} p(s\,|\,n) P_{s \rightarrow t} \, p(m\,|\,t),
\label{ap_chain}
\eeq
where $p(s|n)$ is a conditional probability, defined by $p(n,s)=p(s|n)p(n)$, and $P_{s \rightarrow t}$ is the exact chain (when $p(s|n)=\delta_{E_n,s}$).
Note that $\sum_s p(s|n)=1$ and $\sum_m p(m|t)=1$, so that 
$P_{n \rightarrow m}'$ is a stochastic matrix. 
%BMT
Let us make a few remarks about the behavior of such an approximate 
equilibration process. If this new Markov chain is close to the exact Markov chain, we can 
bound the deviation from the exact fixed point with perturbation theory
\cite{schwei}. Let 
\beq
P_{n \rightarrow m}'=P_{n \rightarrow m}+E_{nm},
\eeq
where $E_{nm}$ is a deviation matrix defined by Eq. (\ref{ap_chain}). 
Let $\rho_{\Delta}=\rho_{s,\beta}'-\rho_{s,\beta}$ where $\rho_{s,\beta}'$ is
the fixed point of the Markov chain $P_{nm}'$. Assume that $P$ is 
diagonalizable. Let $Y$ be the matrix defined as
\beq
Y=({\bf 1}-P+P^{(\infty)})^{-1}-P^{(\infty)},
\eeq
where $P^{(\infty)}$ is the infinite iteration of $P$. We can write 
$P^{(\infty)}=\mbox{diag}(1,0,\ldots,0)$ in the basis where the stationary 
state $\rho_{s,\beta}$ is an eigenvector. In this basis, with 
diagonalizability, $P$ is of the form $\mbox{diag}(1,\lambda_2,\ldots,\lambda_N)$. We can then write 
\beq
Y=\mbox{diag}(0,\frac{1}{1-\kappa},\ldots,\frac{1}{1-\lambda_N}),
\eeq
where $\kappa$ is the second largest eigenvalue. For later use we note that 
the norm $\pr3 Y \pr3_2=\frac{1}{|1-\kappa|}$. It is possible to write the deviation 
$\rho_{\Delta}$ in terms of $Y$ and $E$:
\beq
\rho_{\Delta}= ({\bf 1}-YE)^{-1} Y E \rho_{s,\beta} 
\label{deltaform}
\eeq
when $E$ is small enough such that ${\bf 1}-YE$ is invertible. This expression
can be derived from $P^{(\infty)}\rho_{\Delta}=0$, which follows
from the uniqueness of the stationary state $\rho$. We now use
\beq
\prl \rho_{\Delta} \prl_{tr} \leq \sqrt{N} \prl \rho_{\Delta} \prl_2,
\label{bound2}
\eeq
as in Proposition \ref{unique}. Then using the expression for $Y$, 
Eq. (\ref{deltaform}) and Eq. (\ref{bound2}) (see also below Eq. (\ref{simm})) we can bound 
\beq
\prl \rho_{\Delta} \prl_{tr} \leq C_N\, 
\tr \rho_{s,\beta}^2 
\left(1-\frac{\pr3 E \pr3_2}{|1-\kappa|}\right)^{-1} \frac{\pr3 E \pr3_2}{|1-\kappa|} \leq C_N \left(1-\frac{\pr3 E \pr3_2}{|1-\kappa|}\right)^{-1} \frac{\pr3 E \pr3_2}{|1-\kappa|}.
\eeq
Thus the size of the correction $\rho_{\Delta}$ will be determined by 
the strength of the perturbation $\pr3 E \pr3_2$ and the rate of convergence
of the original Markov chain $P$.
%end BMT

%More general methods exist when $P$ is not diagonalizable and the normal 
%inverse does not exist \cite{schwei} (one uses the Drazin generalized inver%se). 

For a general $H_s$, the computation of an $m$-bit approximation of the 
eigenvalues is likely to cost an exponential (in $m$) number of 
elementary gates. As there are $2^n$ eigenvalues, knowing the $m=\log {\rm poly}(n)$ bits of the 
values of $E_n$ still leaves groups of $\frac{2^n}{{\rm Poly}(n)}$
eigenvalues indistinguishable. Thus only in very special cases, if the gates $U_s^{2^m}$ can 
be implemented with a polynomial (in $m$) number of elementary steps (as 
in Shor's factoring algorithm \cite{shor}) is it possible to compute 
the eigenvalues to high accuracy efficiently.

We have demonstrated a way to set up a Markov chain on a quantum computer 
that will converge to the equilibrium state for long enough time. For 
special Hamiltonians, there might be more efficient ways to tune and modify
this kind of algorithm. The rule ${\cal R}$ might be chosen to depend on 
other features of the eigenstates $\ket{n}$ and $\ket{m}$ as in the classical Metropolis
algorithm where transitions are made between states that are related by
local spin flips. There might be Hamiltonians for which the calculation 
of an eigenvalue, given the eigenvector, is efficient. Then there is 
the hard question of the (rapidly) mixing properties of the chain, that 
determines the computational efficiency of the algorithm.

\section{(Time-dependent) observables}
\label{time}

Given that we have prepared $n$ qubits in the equilibrium state corresponding
to a certain Hamiltonian $H_s$, we can then proceed by experimenting and measuring. The simplest measurement that we could try to perform 
is the estimation of the expectation value of a $c$-local (Hermitian) 
observable $O$:
\beq
\< O \>_s=\tr \rho_{s,\beta}\, O.
\eeq
As $O$ is local, we write $O=\sum_{i=1}^{{\rm poly(n)}} O_i$ where each operator $O_i$ operates on a Hilbert space of constant dimension $c$. We can 
calculate the eigenvectors and eigenvalues of each $O_i$ rapidly on a (possibly)
classical computer, which takes ${\rm poly}(n,c)$ operations.
If $O_i$ has eigenvalues $\mu_i$ that are both smaller as well as 
larger than zero, we define 
$O^{+}_i$ as
\beq
O^+_i=\frac{1}{\max_k \mu_k+|\min_k \mu_k|}(O_i+|\min_k \mu_k|{\bf 1})
\eeq
such that $O^+_i$ is positive semi-definite and has eigenvalues smaller than or 
equal than 1. If $O_i$ has only positive or zero eigenvalues, we just 
``normalize'' the operator by dividing by $\max_k \mu_k$, and similarly
if $O$ has only negative eigenvalues.
Let $I$ be a positive operator valued measurement (POVM \cite{povm}) with operation elements $A_{1,i}$ and $A_{2,i}$
and corresponding outcomes 1 and 2 such that  
\beq
\ba{l}
E_{1,i}=A_{1,i}^{\dagger} A_{1,i}=O^{+}_i, \\
E_{2,i}=A_{2,i}^{\dagger} A_{2,i}={\bf 1}-O^{+}_i.
\ea
\eeq
This measurement will give outcome 1 with probability
\beq
p_{1,i}=\mbox{Tr } O^{+}_i \rho\;\;  {\rm etc}.
\eeq
The operators $A_{1,i}$ and $A_{2,i}$ are given by
\beq
\ba{lr}
A_{1,i}=U_o\, ({\rm diag}_{O^+_i})^{1/2}\, U^{\dagger}_o &\mbox{ and   } A_{2,i}=U_o\, 
({\bf 1}-{\rm diag}_{O^+_i})^{1/2} \,U^{\dagger}_o,
\ea
\eeq
where ${\rm diag}_{O^+_i}$ is the diagonal form of $O^+_i$ and 
$U_o$ the diagonalizing matrix. We summarize these results in a Proposition:

\begin{propo}
The estimation of $\tr \rho \,O$ where $O$ is a $c$-local observable with 
precision $\delta$ and error-probability $\epsilon$ and 
$\rho \in B_{{\rm pos},1}({\cal H}_N)$ ($N=2^n$) takes
$T\,O(\ln \frac{1}{\epsilon}/\delta^2){\rm poly}(n,c)$ operations where $T$ is the time to prepare the state $\rho$.
\label{est_O}
\end{propo}

{\it Proof}:
All commuting observables $O_i$ can be measured once for a single 
preparation of $\rho$. To estimate a probability $p$ with precision 
$\delta$ and error probability $\epsilon$ we need 
$O(\ln \frac{1}{\epsilon}/\delta^2)$ samples \cite{feller}.
$\Box$.

More interesting is an algorithm to estimate time-dependent expectation values.
Let $O_1$ and $O_2$ be two $c$-local observables. We consider how to estimate a time-dependent quantity (identical to Eq. (\ref{corr}))
\beq
\tr \rho_{\beta}\, [O_1, {O_{2}}_t],
\label{2cor}
\eeq
where ${O_{2}}_t$ is in the Heisenberg representation.  Notice that ${O_{2}}_t$, the time-evolved operator, will for general $t$ {\it not} be local. 
Thus we cannot use Proposition \ref{est_O}. The way these quantities come about 
in linear response theory \cite{green_fun} provides the key for how to estimate 
them on a quantum computer. One considers a system that is perturbed at some initial time $t=0$: its time evolution is generated by the perturbed Hamiltonian $H_s+\lambda O_1(t)$ ($O_1(t < 0)=0$) and
$\lambda$ is small. After time $t$ we consider the
response of the system to the perturbation by measuring another observable 
$O_2$. Notice that with Proposition \ref{est_O}, it is simple to perform this 
experiment. Linear response means that we 
take into account corrections of order $\lambda$, but no higher order, 
in the estimation of 
\beq
\delta \<O_2\>_s =\tr O_2 \rho_t -\tr O_2 \rho_{\beta}, 
\eeq
where $\rho_t$ is the time-evolved system density matrix. This first-order 
correction takes the form \cite{fetter} 
\beq
\delta \<O_2 \>_s \approx i\lambda \int_0^t\,dt'\; \tr \rho_{\beta}\, 
[O_1(t'), {O_2}_{t-t'}]. 
\label{lin}
\eeq
If the disturbance $O_1(t)=O_1 \delta(t=0)$ we find on the right hand side 
the correlation function of Eq. (\ref{2cor}). The quantity of 
Eq. (\ref{2cor}) is interesting, because it can be used to compute the simplest reponse of the system, the 
linear response of Eq. (\ref{lin}), which we can directly estimate on our 
quantum computer, provided that both $O_1$ and $O_2$ are local.  
But we are of course not restricted to a linear response regime: 
$\lambda$ is a parameter that we can tune freely. 
A sequence of measurements could determine higher response functions
that will involve quantities such as
\beq
\< {O_1}_{t_1} {O_2}_{t_2} {O_3}_{t_3} \ldots {O_k}_{t_k}\>_s.
\eeq

\section{Conclusion} 
\label{concl}
%BMT

It seems that by asking the question of how fast real quantum systems
equilibrate, we have opened a Pandora's box of hard-to-answer questions.
If there are many simple quantum systems in nature that equilibrate slowly 
(that is, {\em not} in polynomial time) by any dynamics that does not require
extensive preknowledge of the system, then it would be unreasonable 
to ask our quantum computer to perform this task efficiently. By relaxation in
polynomial time we mean the following: in polynomial time we obtain a state
that is within $\epsilon$ trace distance of the equilibrium state where 
$\epsilon$ is a small constant. It might be the case that leaving aside
the classical phenomenon of frustration, relaxation does {\it not} take
place in polynomial time. The idea here is that for a quantum system, the 
eigenbasis is not known beforehand, but must be singled out on the basis 
of an estimation of the eigenvalues, which is generically a hard problem.

This however is not in contradiction with physical and experimental 
reality as we know it, as the quantities that are measured in an 
experimental setup usually involve operators on a small number of qubits; 
these are the experiments that can be done efficiently (in polynomial time) and 
thus do not necessarily probe the system's complete state. For example, the outcomes of the
set of measurements $\sigma_{i_1} \otimes \sigma_{i_2} \otimes \ldots \sigma_{i_n}$ where $\sigma_{i_j}$ is one of the Pauli matrices or ${\bf 1}$,  
completely determines the state, but there are $4^n$ measurements in this
set. In an experimental setup, we might randomly select a polynomial subset 
of them and there is some small chance of order $\frac{poly(n)}{4^n}$ that 
these are the measurements that distinguish the equilibrated state from
the present state in the lab that is supposed to approximate it. 
The estimates of time-dependent correlations could possibly be more 
sensitive to distance from equilibrium, as these involve time-evolved, non-local operations.
The numerical study suggests that product baths whose size is polynomially 
related to the system can function as adequate baths in the sense of 
providing relaxation in polynomial time. The relaxed state could still
be a rather rough approximation to the true equilibrium state, but, 
as we argued above, it might be a good starting point for 
subsequent measurements.

We have taken the bath to be part of the (cost of) the quantum computer. In any experimental setup, there is a natural bath that is used to
equilibrate and cool the quantum computer. Can we use this bath for 
a computational problem such as equilibration?  Consider for example the NMR
quantum computer \cite{nmr} where computation takes place at room 
temperature. In the regime in which the heat bath has a non-Markovian 
character it has been shown to be possible to alter the Hamiltonian of the 
system and the coupling to the bath dynamically (see \cite{lorenza}, but 
also standard books on NMR \cite{abragam}). These techniques could make it 
possible to simulate the time-evolution of a ``designer'' Hamiltonian and 
also to equilibrate the system to the equilibrium state of this designer 
Hamiltonian.

\section{Acknowledgements}
\label{ack}
We would like to thank Charles Bennett, Daniel Loss, 
John Smolin, Ashish Thapliyal, Reinhard Werner, and Ronald de Wolf 
for stimulating discussions. DPD thanks support from the Army Research Office
under contract number DAAG55-98-C-0041.

\appendix 

\section{Norms}
\label{norms}

In this Appendix we give the definitions of several norms and inner products.
The inner product between vectors in ${\bf C}^{N^2}$ can be represented on $B({\cal H}_N)$ as 
\beq
\<\chi_1|\chi_2\>=\tr \chi_1^{\dagger} \chi_2.
\eeq
The trace norm \cite{werner,akn} is defined as
\beq
\prl A\prl_{tr}=\mbox{Tr} \sqrt{A^{\dagger}A}.
\eeq
What makes this norm attractive is that it captures a measurable 
closeness of two density matrices $\rho_1$ and $\rho_2$ \cite{akn}:
\beq
\prl \rho_1-\rho_2\prl_{\rm tr}=\max_A \sum_j |P_1^A(j)-P_2^A(j)|,
\label{close}
\eeq
where $P_1^A$ and $P_2^A$ are the probability distributions over outcomes $j$ 
that are obtained by measuring observable $A$ on $\rho_1$ and $\rho_2$.
The matrix norm $\pr3.\pr3_2$ is defined as
\beq
\pr3 A\!\pr3_2 =\max_{x: \prl x\prl_2=1} \prl\!Ax\!\prl_2. 
\eeq
where $\prl.\prl_2$ is the Euclidean norm on $\dubbelC^{N^2}$: 
$\sqrt{\langle {v}|v \rangle}$ for $\ket{v} \in \dubbelC^{N^2}$. We have
\beq
\prl\!A x\!\prl_2 \leq \pr3 A\!\pr3_2 . \prl\!x\!\prl_2.
\eeq

In order to aid in the interpretation of the numerical results of section \ref{num},  
we present some numerical estimates for the average $\prl.\prl_{tr}$ distance of two randomly chosen density matrices. We first have to choose a measure over $B_{{\rm pos},1}$. All density matrices can be written as $\rho=\sum_i \lambda_i \rho_{ii}$ with 
$\sum_{i=1}^{N} \lambda_i=1$. The eigenvalues $\lambda_k$ lie on a $(N-1)$- 
dimensional simplex $S$ in $\dubbelR^N$. We use the Euclidean metric 
$\prl.\prl_2$ induced on the simplex. The Haar measure on the group
of unitary matrices $U(N)$ induces a uniform measure on the set of 
projectors $\{\rho_{ii}\}_{i=1}^{N^2}$. Together this defines a measure ${\cal M}_{B_{\rm pos,1}}$ \cite{measure}. Within this measure, one can express the average distance between two density matrices $\rho_1$ and $\rho_2$, using the unitary invariance of $\prl.\prl_{tr}$, as
\beq
\ba{c}
[\prl \rho_1-\rho_2\prl_{tr}]_{{\cal M}_{B_{\rm pos,1}}}= 
\frac{1}{{\rm Vol}(S)^2 V(U(N))}
\int dU \int_0^1 d\lambda_1 \ldots d\lambda_k \int_0^1 d\mu_1 \ldots d\mu_k \;\delta(\sum_i \lambda_i-1)\, \delta(\sum_i \mu_i-1) \\

\mbox{Tr} |\sum_j \lambda_j \rho_{jj}- U \sum_j \mu_j \rho_{jj} U^{\dagger}|.
\ea
\label{av_dist}
\eeq

The values obtained by a numerical calculation of Eq. (\ref{av_dist}) are 
tabulated in Table \ref{gram}.

\section{Preparation of the bath}
\label{bath}

To prepare the state 
\beq
\rho_{b,\beta}=\rho^1_{b,\beta} \otimes \ldots \otimes \rho^k_{b,\beta}, 
\eeq
given $H_b=\sum_{i=1}^{k} {\bf 1}_{K/2} \otimes h_i$, we first calculate 
the eigenvalues and eigenvectors of each qubit Hamiltonian $h_i$. We prepare the state
\beq
\Pi_{i=1}^k \left( e^{-\beta e_{i,0}} \ket{0}\bra{0}+e^{-\beta e_{i,1}}\ket{1}\bra{1}\right)/Z_i.
\eeq
with $\{e_{0,i},e_{1,i}\}$ the eigenvalues of qubit Hamiltonian $h_i$. This can be done by changing an initial state $\ket{0}\bra{0}$ with probability 
$e^{-\beta e_{i,1}}/Z_i$ into state $\ket{1}\bra{1}$ for each $i$.
We then rotate each 
qubit to its eigenbasis $\{\ket{{b_i}_0},\ket{{b_i}_1}\}$:
\beq
\otimes_{i=1}^k U_{b_i}=\otimes_{i=1}^k (\ket{{b_i}_0}\bra{0}+\ket{{b_i}_1}\bra{1}).
\eeq

In total we perform $2k$ elementary qubit operations plus some constant
classical overhead.

%%%%%%%%%%%%%%%%%%%%%%%%%%%%%%%%%%%%%%%%%%%%%%%%%%%%%%%%%%%%%%%%%%%%%%%%%%%%%

% \end{multicols}

\newpage

\begin{table}[p]
\centering
\begin{tabular}{l|c|c|c|r}
      & $H_s$ & $H_b$ & $S$ & $B$  \\
\hline
dimension  & $N=2,..,2^4$ & $K=2^2,..,2^6$ & N  & K  \\
\hline
locality  & $c_s=4$ & $c_b=2$ & 4 & 4 \\
\hline
sampling scale a & 1 & $f(n,k,c_s,c_b)$ & 1 & 1 \\
\end{tabular}
\caption{Some settings in the numerical simulation.}
\label{settings}
\end{table}

\newpage

\begin{figure}[p]
\centering
\epsfxsize=8.0cm
\epsfbox{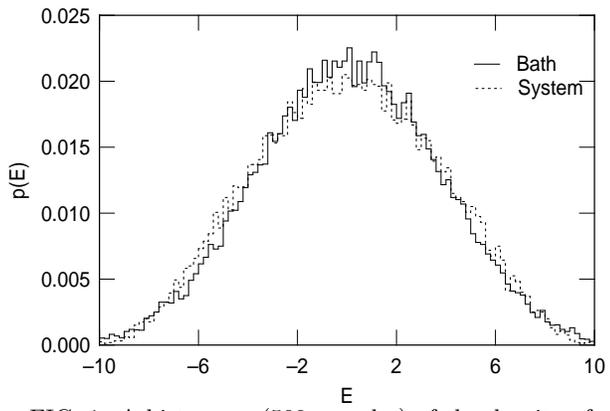}
\caption{A histogram (500 samples) of the density of states (unnormalized) for $N=32$ and $K=64$ with 
sampling scale set as Eq. (\ref{ss}).}
\label{dens}
\end{figure}

\newpage

\begin{figure}[p]
\epsfxsize=16.0cm
\epsfbox{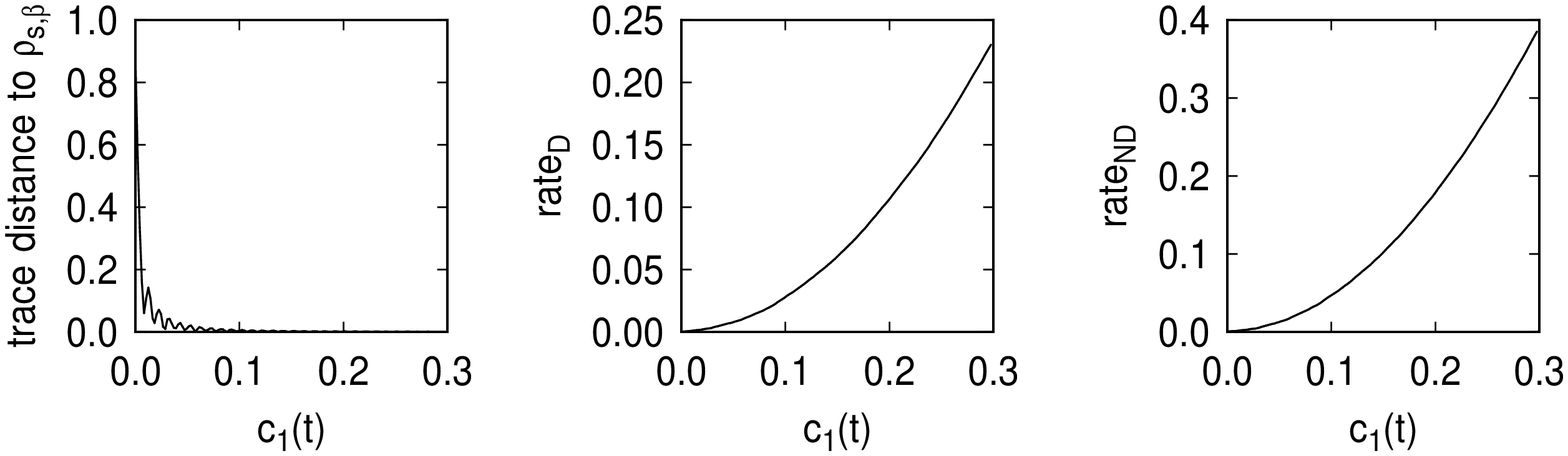}
\caption{An example of succesful equilibration for $n=1$, $k=3$ and $\beta=3$.}
\label{fig_yes}
\end{figure}

\newpage

\begin{figure}[p]
\epsfxsize=16.0cm
\epsfbox{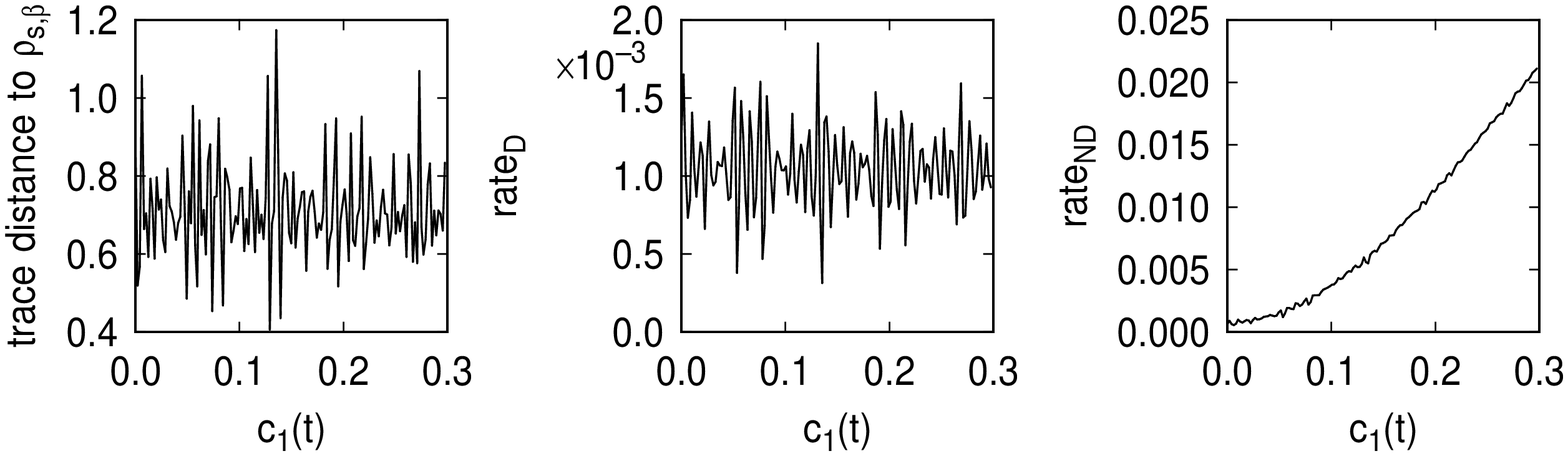}
\caption{An example of an unsuccesful equilibration for $n=1$, $k=3$ and $\beta=3$.}
\label{fig_no}
\end{figure}

\newpage

\begin{figure}[p]
\centering
\epsfxsize=16.0cm
\epsfbox{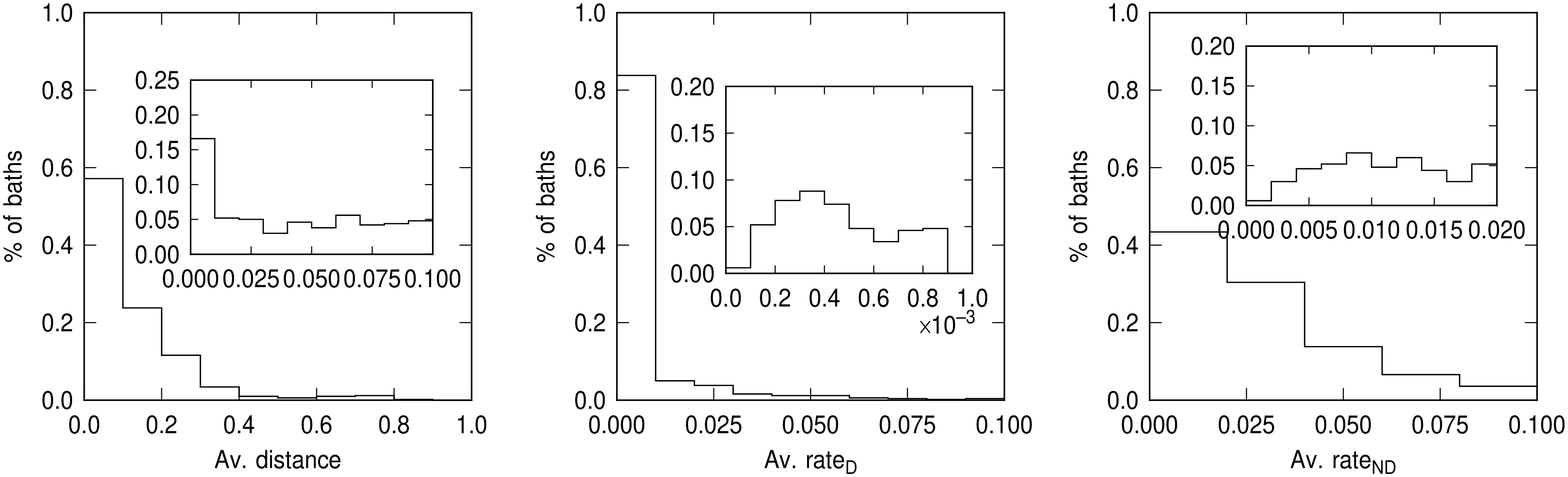}
\caption{An example of the distribution of baths (500 samples) for $n=2$ and $k=3$ and 
$\beta=2$.}
\label{histo}
\end{figure}

\newpage

\begin{figure}[p]
\epsfxsize=16.0cm
\epsfbox{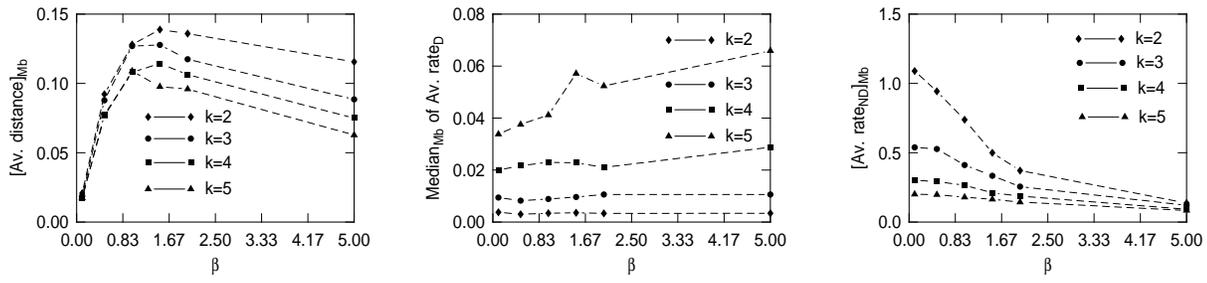}
\caption{Means and median for n=1 (500 samples).}
\label{dim2}
\end{figure}

\newpage

\begin{figure}[p]
\epsfxsize=16.0cm
\epsfbox{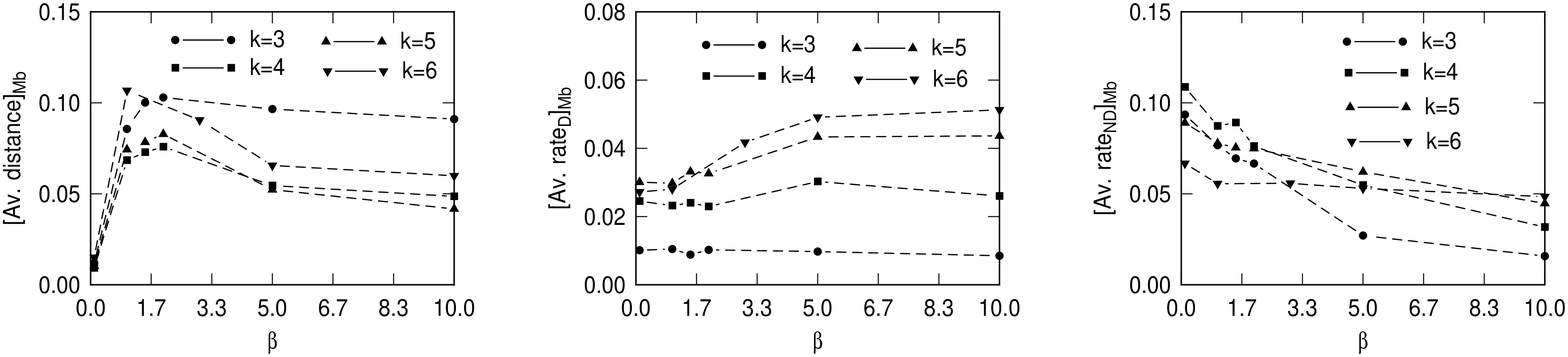}
\caption{Means for n=2 (200-500 samples).}
\label{dim4}
\end{figure}

\newpage

\begin{figure}[p]
\epsfxsize=16.0cm
\epsfbox{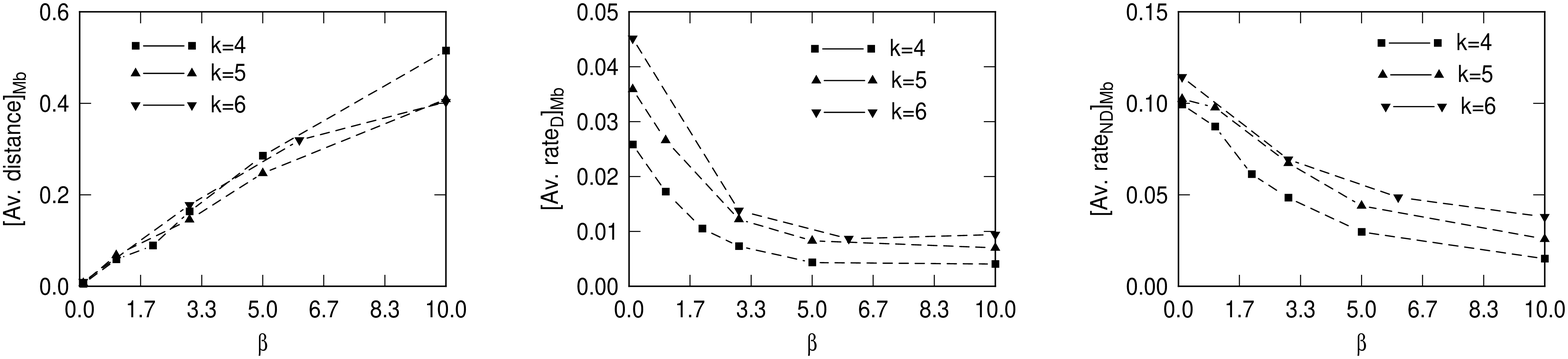}
\caption{Means for n=3 (50-100 samples).}
\label{dim8}
\end{figure}

\newpage

\begin{figure}[p]
\epsfxsize=16.0cm
\epsfbox{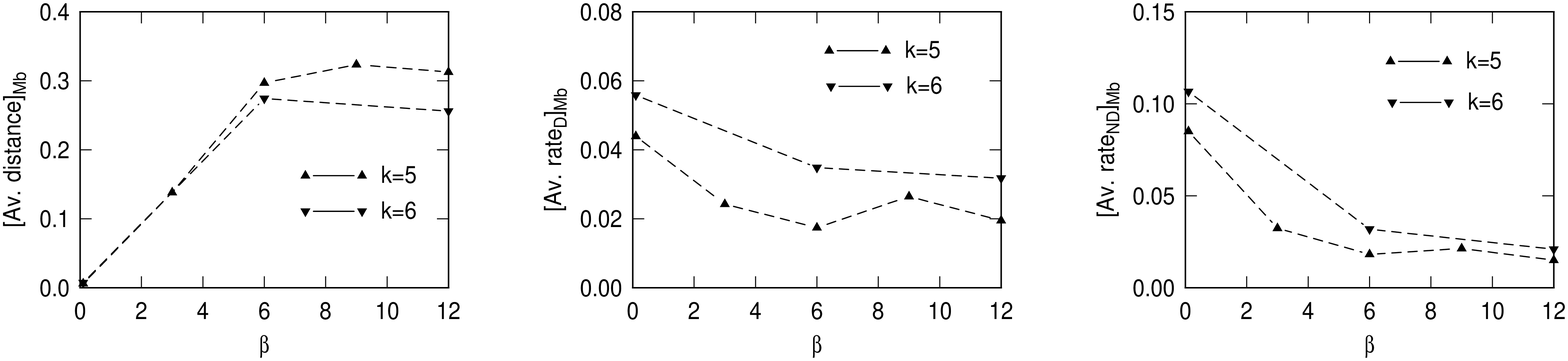}
\caption{Means for n=4 (15-20 samples).}
\label{dim16}
\end{figure}

\newpage

\begin{table}[p]
\centering
\begin{tabular}{l|c|r}
dim N & mean & standard error$=\sqrt{\mbox{var}/(n-1)}$, $n=1000$ \\
\hline
4 &  0.90388 & 0.00740588 \\
\hline
8 &  0.96190 & 0.00514057   \\
\hline
16 & 1.00294 &  0.00341226 \\
\hline
32 & 1.01452  & 0.00220363 \\
\hline
64 & 1.02617 &  0.00132233 \\
\end{tabular}
\caption{The average distance between two randomly selected density matrices.}
\label{gram}
\end{table}


\begin{thebibliography}{99}
\bibitem{ofer_com}In \cite{ofer} quantum algorithms were presented to sample
from the equilibrium distribution of arbitrary Ising spin glass models. The 
performance of these algorithms does not necessarily provide an exponential 
speedup over the best classical algorithms (see \cite{jerrum}) for the hard 
instances though.
\end{thebibliography}
\end{document}